%% file: Bayesian_Clustering_of_Multiple_Zero-Inflated_Outcomes.tex
\documentclass[11pt]{article}
\usepackage[utf8]{inputenc}
\usepackage[a4paper, total={5.5in, 10in}]{geometry}

\usepackage{authblk}
\usepackage{amsmath}
\usepackage{graphicx}

\usepackage{psfrag,epsf}
\usepackage{enumerate}
\usepackage{url} 

\usepackage{nicefrac}
\usepackage{float}
\usepackage{amssymb,verbatim}
\usepackage{bm,lscape}
\usepackage{wasysym}
\usepackage{bbm}
\usepackage{bigints}
\RequirePackage[colorlinks,citecolor=blue,urlcolor=blue]{hyperref}

\usepackage{theorem}
\usepackage{subcaption}
\usepackage{algorithm}
\usepackage{algpseudocode}

\usepackage{graphicx,wrapfig}
\usepackage{tikz}
\usetikzlibrary{fadings}
\usetikzlibrary{matrix}
\usetikzlibrary{calc}
\tikzset{
	zero/.style = {minimum width=.5cm, minimum height=.5cm, draw, fill={rgb,255:red,255; green,102; blue,102}},
	one/.style = {zero, fill={rgb,255:red,102; green,178; blue,255}},
	titlerect/.style={inner ysep=2ex,inner
		xsep=1ex,minimum width={2*width("#1")},align=center,label={[anchor=center,fill=white,font=\Large\bfseries\sffamily]above:#1}},
}

\allowdisplaybreaks

\def\bSig\mathbf{\Sigma}

\definecolor{darkblue}{rgb}{0.0, 0.0, 0.55} 
\definecolor{dgreen}{rgb}{0.0, 0.5, 0.0}
\definecolor{blue-violet}{rgb}{0.54, 0.17, 0.89}

\newcommand{\beq}{\begin{equation}}
	\newcommand{\eeq}{\end{equation}}
\newcommand{\iid}{\overset{\textnormal{iid}}{\sim}}

\allowdisplaybreaks

\title{Bayesian clustering of multiple zero-inflated outcomes}

\author[1]{Beatrice Franzolini}
\author[2]{Andrea Cremaschi}
\author[3]{Willem van den Boom}
\author[4]{Maria De Iorio}

\affil[1,2,4]{Singapore Institute for Clinical Sciences, 
	 Agency for Science, Technology and Research}
\affil[3,4]{Yong Loo Lin School of Medicine, National University of Singapore}
\affil[4]{Department of Statistical Science, University College London}
\date{ }

\usepackage{natbib}

\begin{document}
\maketitle

\begin{abstract}  
	Several applications involving counts present a large proportion of zeros (excess-of-zeros data). A popular model for such data is the Hurdle model, which explicitly models the probability of a zero count, while assuming a sampling distribution on the positive integers. We consider data from multiple count processes. In this context, it is of interest to study the patterns of counts and cluster the subjects accordingly.
	We introduce a novel Bayesian nonparametric approach to cluster multiple, possibly related, zero-inflated processes. We propose a joint model for zero-inflated counts, specifying a Hurdle model for each process with a shifted Negative Binomial sampling distribution. Conditionally on the model parameters, the different processes are assumed independent, leading to a substantial reduction in the number of parameters as compared to traditional multivariate approaches. The subject-specific probabilities of zero-inflation and the parameters of the sampling distribution are flexibly modelled via an \emph{enriched} finite mixture with random number of components. This induces a two-level clustering of the subjects based on the zero/non-zero patterns (outer clustering) and on the sampling distribution (inner clustering). Posterior inference is performed through tailored MCMC schemes. 
	We demonstrate the proposed approach on an application involving the use of the messaging service WhatsApp.
\end{abstract}


\maketitle

\section{Introduction}

Count data presenting excess of zeros are commonly encountered in applications. These can arise in several settings, such as healthcare, medicine, or sociology. In this scenario, the observations carry structural information about the data-generating process, i.e., an \emph{inflation} of zeros.
The analysis of zero-inflated data requires the specification of models beyond standard count distributions, such as Poisson or Negative Binomial. Commonly used models are the Zero-Inflated \citep{lambert1992zero}, the hurdle \citep{mullahy1986specification} and the zero-altered \citep{heilbron1994zero} models. The first class assumes the existence of a probability mass at zero and a distribution over $\mathbb{N}_0 = \{0, 1, 2, \dots\}$. This type of models explicitly differentiates between the zeros originating from a common underlying process, such as the utilisation of a service, described by the sampling distribution on $\mathbb{N}_0$, and those arising from a structural phenomenon, such as the ineligibility to use the service, which are modelled by the point mass. 
Very popular Zero-Inflated models are the Zero-Inflated Poisson (ZIP) and the Zero-Inflated Negative Binomial (ZINB) models, where the sampling distribution is chosen to be a Poisson and a Negative Binomial, respectively. These models allow for inflation in the number of zeros and departures from standard distributional assumptions on the moments of the sampling distribution. For instance, the ZIP model allows the mean and the variance of the distribution to be different from each other (as opposed to a standard Poisson distribution), while the ZINB additionally captures overdispersion in the data.

Hurdle models are a very popular choice of distributions for modelling zero-inflated counts. Differently from the Zero-Inflated ones, these models handle zeros and positive observations separately, assuming on the latter a sampling distribution with support on $\mathbb{N} = \mathbb{N}_0 \setminus \{0\}$. Thus the distribution of the count data is given by: 
\begin{equation}
	\mathbb{P}\left(Y_i = y_i\right) =
	\begin{cases}
		\left(1 - p_i\right), & y_i = 0 \\
		p_i g\left(y_i \mid \bm \mu_i\right), & y_i > 0
	\end{cases}
\end{equation}
where $p_i$ and $g$ now capture two distinct features of the data. Hurdle models present appealing features that can make them preferable to Zero-Inflated models. Firstly, hurdle distributions allow for both inflation and deflation of zero-counts. Indeed, under a Zero-Inflated model, the probability of observing a zero is always greater than the corresponding probability under the sampling distribution, thus making it impossible to capture deflation in the number of zeros \citep{min2005random}.
Secondly, and more importantly for our work, the probability of zero counts in hurdle models is independent of the parameters controlling the distribution of non-zero counts. This feature improves interpretability and facilitates parameter estimation. Note that the Zero-Altered model proposed by \citet{heilbron1994zero} is a modified hurdle model in which the two parts are connected by specifying a direct link between the model parameters.

Univariate models for zero-inflated data can be extended to multivariate settings, where several variables presenting excess of zeros are recorded, e.g.\ in applications involving questionnaires or microbiome data analysis. In this context, a multivariate extension of the ZIP model has been proposed by \citet{li1999multivariate}, through a finite mixture with ZIP marginals. In this construction, the number of parameters increases linearly as the number $d$ of zero-inflated processes increases, as the total number of parameters is $3d +2$. See also \cite{liu2015type}, \cite{liu2019new} and, \citet{tian2018type} for simplified versions of the previous construction involving a smaller number of parameters and better distributional properties.

In a Bayesian parametric setting, \citet{fox2013multivariate} proposes the joint modelling of two related zero-inflated outcomes. Their strategy is based on the ZIP model,  
with the same Bernoulli component to capture the extra zeros for both processes. Correlation between subject-specific outcomes is accounted for through the specification of a joint random effect distribution for the parameters governing the sampling  distribution of the two processes.   Alternatively, \citet{lee2020bayesian} model the binary variables indicating whether an observation is positive or not via a multivariate probit model \citep{chib1998analysis,garcia2007conditional}. In this approach, the vectors of latent continuous variables characterising the multivariate probit are modelled jointly assuming a random unstructured correlation matrix describing their dependence. 

In several applications, knowledge relative to the grouping of the subjects is also available, thus providing additional information that can be exploited in the model \citep{choo2018bayesian}. Moreover, the clustering structure can be estimated by assuming a prior distribution on the partition of the subjects, e.g.\ via the popular Dirichlet process \citep{li2017bayesian} or a mixture with random number of components as proposed in \citet{hu2020zero}. 

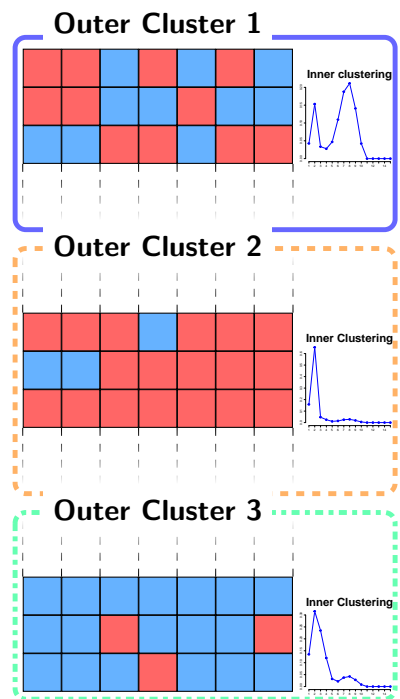
\begin{wrapfigure}{r}{0.48\textwidth}
	\centering
	\input{figures/Figure_ZeroInflated.tikz}
	\caption{Example of two-level clustering induced by the enriched mixture with random number of components. The observations are first clustered based on their zero/non-zero patterns, indicated in the figure in blue and red, respectively. Within each outer cluster, subjects are grouped based on the sampling distribution of the non-zero observations. The inner clustering structure is here depicted via a multimodal discrete distribution, representing a finite mixture.}
	\label{fig:Enriched_Fig}
\end{wrapfigure}

In the context of Bayesian semiparametric approaches, \cite{shuler2021bayesian} propose to model multivariate zero-inflated count data by linking different Dirichlet Process mixtures of ZINB models through the use of the popular dependent Dirichlet process \citep {maceachern1999dependent}. In particular, the probability of zeros and the sampling distribution are modelled via two distinct single-p DDP, where the location parameters of the mixture depend on a categorical covariate. The proposed approach yields flexible estimation of the partition of the subjects, although it does not allow for sharing of information a priori between the two components of the ZINB model, thus yielding two separate clustering structures. A different semiparametric approach is proposed by \citet{arab2012semiparametric}, which exploits the multivariate ZIP construction of \citet{li1999multivariate} to model bivariate count data, but  the proportion of zeros and the intensity of the sampling distribution are modelled through the introduction of spline regression terms. The spline approach is flexible and computationally tractable when $d$ is small. For larger dimensions, this model would induce a non-trivial computational burden.

The focus of this work is clustering of individuals based on multiple, possibly related, zero-inflated processes. To this end, we propose a Bayesian  approach for joint modelling of zero-inflated count data, based on finite mixtures with random number of components. In particular, we specify a hurdle model for each process with a shifted Negative Binomial sampling distribution on the positive integers. Let $n$ denote the sample size and $d$ the number of processes under study. The subject-specific probabilities of zero-inflation $p_{ij}$ for the $i$-th individual and the $j$-th process, $i=1,\ldots,n$, $j=1,\ldots,d$, and the parameter vector of the sampling distribution $\bm \mu_{ij}$ are flexibly modelled via an {\em enriched} mixture with random number of components, borrowing ideas from the Bayesian nonparametric literature on the Dirichlet process. One of the main novelties of our work is to combine a recent representation of finite mixture models with random number of components presented in \citet{Argiento2022} with a finite extension of the enriched nonparametric prior proposed by \citet{wade2011enriched} to achieve a two-level clustering of the subjects, where at the {\em outer} level individuals are clustered based on the pattern of zero/non-zero observations, while within each outer cluster they are grouped at a finer level (which we refer to as {\em inner} level) according to the distribution of the non-zero counts. Figure \ref{fig:Enriched_Fig} provides an illustration of the nested clustering structure.

Enriched priors in Bayesian nonparametrics generalise concepts developed by \citet{consonni2001conditionally}, who propose a general methodology for the construction of enriched conjugate families for the parametric natural exponential families. 
The idea underlying this approach is to decompose the joint prior distribution for a vector of parameters indexing a multivariate exponential family into tractable conditional distributions. In particular, distributions belonging to the multivariate natural exponential family satisfy the  \textit{conditional reducibility} property, which allows reparameterising the distribution in terms
of a parameter vector, whose components are variation and likelihood independent. Then, it is possible to construct an enriched standard conjugate family on the parameter vector, closed under i.i.d.\ sampling, which leads to the breaking down of the global inference procedure into several independent subcomponents. Such parameterisation achieves  greater flexibility in prior specification relative to the standard conjugate one, while still allowing for efficient computations \citep[see, for example,][]{consonni2004reference}. An example of this class of parametric priors is the enriched Dirichlet distribution \citep{connor1969concepts}. 

In a Bayesian nonparametric framework, \citet{wade2011enriched} first propose an enrichment of the Dirichlet process \citep{ferguson1973bayesian} that is more flexible with respect to the precision parameter but still conjugate, by defining a joint  random probability measure on the measurable product space $\left(\mathcal{X},\mathcal{Y}\right)$ in terms of the marginal and conditional distributions, $P_X$ and $P_{Y\mid X}$, and assigning independent Dirichlet process priors to each of these terms. The enriched Dirichlet process enables a nested clustering structure that is particularly appealing in our setting and allows for a finer control of the dependence structure between $X$ and $Y$. This construction has been employed also in nonparametric regression problems to model the joint distribution of the response and the covariates \citep{wade2014improving, Gadd2020}, as well as in longitudinal data analysis \citep{zeldow2021functional} and causal inference \citep{roy2018bayesian}. Recently, \citet{rigon2020enriched} propose the enriched Pitman–Yor process which leads to a more robust clustering estimation.

In this work, we consider the joint distribution of $d$ zero-inflated process, where the $d$-dimensional vectors of probabilities $\left(p_{i1},\ldots,p_{id}\right)$ correspond to $X$, while the parameters of the sampling distributions $\bm \mu_{ij}$ correspond $Y$.  
The enrichment of the prior is achieved by modelling both $P_X$ and $P_{Y\mid X}$  through a mixture with random number of components \citep[see, for instance,][]{miller2018mixture}. We exploit the recent construction by \citet{Argiento2022} based on Normalised Independent Finite Point Processes, which allows for a wider choice of prior distributions for the unnormalised weights of the mixture. Therefore, the proposed model offers more flexibility, while preserving computational tractability. 

The motivating application for the proposed model is the analysis of multiple count data collected from a questionnaire on the frequency of use of the messaging service WhatsApp \citep{clinicaltrials2021}. In particular, the questionnaire concerns the sharing of COVID-19-related information via WhatsApp messages, either directly or by forwarding, over the course of a week. For each subject, responses to the same seven questions are recorded over seven consecutive days, providing information on a  subject's WhatsApp use (see 
Table~S1 in Supplementary Material). In this set-up, the multiple count processes correspond to the seven questions, all of which displaying an excess of zeros (see Figure~S2 in Supplementary Material).

The manuscript is organised as follows. Section~\ref{sec:Model} introduces a novel enriched prior process for multiple zero-inflated outcomes, while Section~\ref{sec:Inference} describes the Markov chain Monte Carlo (MCMC) algorithm designed for posterior inference. We demonstrate the model on  the WhatsApp application in Section~\ref{sec:Application}. We conclude the paper in  Section~\ref{sec:Concludes}.

\section{The model}\label{sec:Model}

\subsection{Likelihood}

Let $Y_{ij}$ be the count of subject $i = 1, \dots, n$ for outcome $j = 1, \dots, d$ and let $\bm Y_i =(Y_{i1},\ldots, Y_{id})$ be the $d$-dimensional vector of observations for subject $i$. To take into account the zero-inflated nature of the data, we assume for each outcome $j$ a hurdle model. Each observed count $Y_{ij}$ is equal to zero with probability $1 - p_{ij}$, while with probability $p_{ij}$ it is distributed according to a probability mass function (pmf) $g\left(\cdot\mid \bm \mu_{ij}\right)$ with support on $\mathbb{N}$. Assuming conditional independence among responses, the likelihood for a subject is given by:
\begin{equation}\label{eq:kernel}
	\begin{aligned}
		\mathbb{P}\left( \bm Y_i = \bm y_i \mid \bm{p}_i, \bm{\mu}_i \right) = \prod\limits_{j=1}^d
		f\left(y_{ij}\mid p_{ij},\bm \mu_{ij}\right)
		\qquad
		f\left(y\mid p,\bm{\mu}\right) = \begin{cases}
			1 - p, &y = 0\\
			p\, g\left(y\mid \bm{\mu}\right), &y>0
		\end{cases}
	\end{aligned}
\end{equation}
with $\bm p_i = \left(p_{i1}, \ldots, p_{id}\right) \in (0,1)^{d}$, $\bm \mu_i = \left(\bm \mu_{i1},\ldots,\bm \mu_{id}\right)$, $i = 1, \dots, n$. In what follows, we set $g$ to be a shifted Negative Binomial distribution with parameters $\bm \mu_{ij} =(r_{ij}, \theta_{ij})$ and pmf: 
\begin{equation}\label{eq:nonzeropart}
	g\left(y \mid r_{ij}, \theta_{ij} \right) = \frac{\left(y + r_{ij} - 2\right)!}{\left(r_{ij} - 1\right)!\left(y - 1\right)!} \theta_{ij} ^{y - 1} \left(1 - \theta_{ij}\right)^{r_{ij}}, \qquad y \in \mathbb{N}
\end{equation}
where $r_{ij} \in \mathbb{N}$ and $\theta_{ij} \in (0,1)$, for $i = 1, \dots, n$ and $j = 1, \dots, d$. Different parametric choices for $g$ are possible (e.g. a shifted Poisson), or even nonparametric alternatives could be employed. Note that the conditional independence assumption among the multiple processes leads to a significant reduction in the number of parameters as compared to multivariate zero-inflated models.

\subsection{Enriched finite mixture model}

In this work, we propose an \textit{enriched} extension of the Normalised Independent Finite Point Process (Norm-IFPP) of \citet{Argiento2022} and specify a joint prior for $\left(\bm p_i, \bm\mu_i\right)$ as conditionally dependent processes. This allows us to account for interindividual heterogeneity, overdispersion, and outliers and induces data-driven nested clustering of the observations. Each subject is first assigned to an \textit{outer} cluster, and then  clustered again at an \textit{inner} level, providing increased interpretability.
Differently from previous work on Bayesian nonparametric enriched processes, we opt for a finite mixture with random number of components, where the weights are obtained through the normalisation of a finite point process. Finite mixture models with random number of components have received increasing attention in the last years \citep[see, for example,][]{ malsiner2016model,miller2018mixture}. The representation of \citet{Argiento2022} allows for the specification of a wide range of distributions for the weights and simultaneously leads to effective and widely applicable MCMC schemes on which   Algorithms~\ref{algorithm} and \ref{alg:marg} are based. More specifically, they show that a finite mixture model is equivalent to a realisation of a stochastic process with random dimension and infinite-dimensional support, leading to flexible distributions for the weights of the mixture given by the normalisation of a finite point process. We thus employ this approach as it allows for efficient computations via a conditional algorithm, as compared to labour-intensive reversible jump algorithms common in mixture models. An alternative efficient conditional sampler for mixtures with a random number of components is the recently proposed telescopic sampler by \cite{fruhwirth2021generalized}.

In the proposed framework, the observations are assumed to be sampled from a mixture with an inner and an outer component. As kernel of the mixture, we assume the hurdle model in \eqref{eq:kernel}, which distinguishes between the probabilities of being non-zero $\bm{p}_i$ and the parameters of the sampling distribution $(\bm{r}_i, \bm{\theta}_i)$. The components of the outer mixture are determined by different probabilities of non-zero outcomes, denoted with $\bm {p_m^\star} = (p^\star_{m1},\ldots, p^\star_{md})$, for $m=1,\ldots,M$, with $M$ the number of outer mixture components. The components of the inner mixtures are characterised by distinct parameters of the sampling distribution, denoted with $\bm {r^\star_{ms}} = (r^\star_{ms1},\ldots, r^\star_{msd})$ and $\bm \theta^\star_{ms} = (\theta^\star_{ms1},\ldots, \theta^\star_{msd})$, for $s=1,\ldots,S_m$ and $m=1,\ldots,M$, where $S_m$ is the number of mixture components within the $m$-th outer mixture component. 
Letting $\bm \psi^\star_{msj} = \left(p^\star_{mj}, r^\star_{msj}, \theta^\star_{msj} \right)$ and $\bm \psi^\star_{ms} = \left(\bm \psi^\star_{ms1}, \dots, \bm \psi^\star_{msd}\right)$, the mixture model is as follows:

\begin{equation}\label{eq:model_general}
\begin{aligned}
	\bm Y_i\mid \{\bm\psi^\star_{ms}\}, \bm w, \{\bm q_{m}\}  & \iid \underbrace{\sum_{m=1}^M w_m}_{\text{outer level}} \quad \underbrace{\sum_{s=1}^{S_m} q_{ms}}_{\text{inner level}} \quad \prod_{j=1}^d f\left(y_{ij} \mid \bm \psi^\star_{msj}\right) \\
	\bm q_m = \left(q_{m1},\ldots,q_{mS_m}\right) \mid S_m & \sim
	\text{Dirichlet}_{S_m}\left(\gamma_S, \ldots, \gamma_S\right)\\
	\bm w = \left(w_1,\ldots,w_M\right)\mid M &\sim \text{Dirichlet}_{M}\left(\gamma_M, \ldots, \gamma_M\right) \\
	\bm p^\star_m &\iid \prod_{j=1}^d \text{Beta}\left(\alpha, \beta\right)\\
	\bm r^\star_{ms} &\iid \prod_{j=1}^d \text{Geometric}\left(\zeta\right)\\
	\bm\theta^\star_{ms} &\iid \prod_{j=1}^d \text{Beta}\left(\eta,\lambda\right)\\
	S_1,\ldots, S_M\mid M  
	&\iid \text{Poi}_0\left(\Lambda_S\right)\\
	M 
	&\sim \text{Poi}_0\left(\Lambda_M\right) 
\end{aligned}
\end{equation}
where the kernel $f\left(y_{ij}\mid \bm \psi^\star_{msj} \right) $ is defined via conditionally independent hurdle models in \eqref{eq:kernel}--\eqref{eq:nonzeropart}. Here $\text{Dirichlet}_{M}(\gamma_M, \ldots, \gamma_M)$ denotes the symmetric Dirichlet distribution defined on the $(M-1)$-dimensional simplex with mean $1/M$, which is the distribution of the normalised mixture weights. $\text{Beta}\left(\alpha, \beta\right)$ indicates the Beta distribution with mean $\alpha/(\alpha + \beta)$ and variance $\alpha \beta / ((\alpha + \beta)^2(\alpha + \beta + 1))$,  $\text{Geometric}\left(\zeta\right)$ the Geometric distribution with mean $1/\zeta$, and $\text{Poi}_0(\Lambda)$ the shifted Poisson distribution, such that if $X \sim \text{Poi}_0(\Lambda)$ then $X - 1$ has a Poisson distribution with mean $\Lambda$. Moreover, $M$ and $S_m$, for $m=1,\ldots,M$, indicate the random number of components at the outer and inner level of the enriched Norm-IFPP, respectively.

The outer mixture is a mixture of multivariate Bernoulli distributions, and coincides with the widely-used Latent Class model \citep{lazzaro}. Moreover, being conditionally independent of the actual values of the non-zero observations, it offers further computation advantages as shown in Section~\ref{sec:Inference}. 

Model~\eqref{eq:model_general} induces a partition of the subject indices $\left\{ 1, \dots, n \right\}$ at an outer and an inner level. Let  $c_i$ and $z_i$, for $i=1,\ldots,n$, denote the allocation variables which  indicate to which component of the mixture each subject is assigned to at the outer and inner level, respectively.  When two subjects, $i$ and $l$, are assigned to the same component of the outer level mixture, then the probabilities of observing a zero for the two subjects are the same, $\bm p_i = \bm p_l$, and the two subjects are assigned to the same cluster, i.e.\ $c_i=c_l$. Moreover, if the two subjects are also assigned to the same component of the inner level mixture, we have $z_i=z_l$ and $\bm \mu_i = \bm \mu_l$ (with obviously $c_i=c_l$). However, the vectors of parameters $\bm \mu_i$ and $\bm \mu_l$  characterising the sampling distribution might be different even when $c_i=c_l$ and, consequently, the two subjects might be assigned to different clusters at the inner level. This is reflected in the components of the vectors of parameters $\left(\bm p_i, \bm \mu_i\right)$ and $\left(\bm p_l, \bm \mu_l\right)$, which might share only the component corresponding to the probability of zero outcomes or both components. 

Using allocation variables, the conditional dependence structure between outer and inner levels is the following. Let 
\begin{equation}
\label{eq:ytilde}
\widetilde{Y}_{ij} =
\begin{cases}
	1 &\text{if } Y_{ij} > 0\\
	0 &\text{if } Y_{ij} = 0
\end{cases}
\end{equation}
$\widetilde{\bm Y}_i = \left( \widetilde{Y}_{i1},\ldots,  \widetilde{Y}_{id}\right)$, $\mathcal{C}_m = \{i: c_i=m\}$, and $\mathcal{C}_{ms} = \{i: c_i=m, z_i = s\}$.
\noindent
\newline Outer mixture:
\begin{equation}
\label{eq:outermix}
\begin{aligned}
	\widetilde{\bm Y}_i \mid \bm p_i & \sim   \prod_{j=1}^d p_{ij}^{\widetilde{y}_{ij}}\left(1- p_{ij}\right)^{\widetilde{y}_{ij}}, \qquad \widetilde{y}_{ij} \in \{0,1\}
	\\
	\bm p_i & = \bm p^\star_{c_i} \\
	\bm p^\star_{1},\ldots, \bm p^\star_{M} \mid M &\iid \prod_{j=1}^d \text{Beta}\left(\alpha , \beta\right) \\ 
	\Pr\left(c_i = m\right) & \propto \Gamma_m , \quad m=1,\ldots ,M\\
	\Gamma_1, \ldots, \Gamma_M & \iid \text{Gamma}\left(\gamma_M,1\right) \\ 
	M 
	&\sim \text{Poi}_0\left(\Lambda_M\right) 
\end{aligned}
\end{equation}
\newline \noindent Inner mixture:
\begin{align*}
\bm Y_i \mid M, c_i = m, \bm p^\star_m, \bm r_{mi}, \bm \theta_{mi}   & \sim \prod_{j=1}^d f\left(y_{ij} \mid p_{mj}^\star , r_{mij}, \theta_{mij}\right)\\ 
\left(\bm r_{mi},\bm \theta_{mi}\right) & = \left(\bm r_{mz_i}^\star,\bm \theta_{mz_i}^\star\right) \\
\bm r_{m1}^\star,\ldots ,   \bm r_{mS_m}^\star \mid S_m & \iid \prod_{j=1}^d \text{Geometric}\left(\zeta\right) \\
\bm \theta_{m1}^\star,\ldots ,   \bm \theta_{mS_m}^\star \mid S_m & \iid \prod_{j=1}^d \text{Beta}\left(\eta,\lambda\right)\stepcounter{equation}\tag{\theequation}\label{eq:innerrmix} \\
\Pr\left(z_i = s\mid c_i= m\right) & \propto \Delta_{ms} , \quad i\in \mathcal{C}_m,\quad s=1,\ldots , S_m \\
\Delta_{m1}, \ldots, \Delta_{mS_m} & \iid \text{Gamma}\left(\gamma_S,1\right) \\S_1,\ldots, S_M \mid M 
&\iid \text{Poi}_0\left(\Lambda_S\right)
\end{align*}
where, as before, we denote with $\bm{p_m}^{\star}$, $\bm{r_{ms}}^{\star}$ and $\bm{\theta_{ms}}^{\star}$ the component-specific parameters, which are assumed a priori independent and $\text{Gamma}\left(\alpha,\beta\right)$ is the Gamma distribution with mean $\alpha/\beta$ and variance $\alpha/\beta^2$. The choice of Gamma distribution for the unnormalised weight of the mixture  leads to the standard Dirichlet distribution for the normalised weights. In this setting, the computations are greatly simplified by the introduction of a latent variable, conditionally on which the unnormalised weights are independent. See \citet{Argiento2022} for details. Note that the inner mixture is here defined conditionally on the probabilities $p_{m,j}$ of being zero and not on $\tilde{\bm{Y}}_{i}$. Thus, while conditioning on $p_{m,j}$, $\bm{Y}_i$ is still allowed to present zero entries. Finally, we highlight that representations \eqref{eq:model_general} and \eqref{eq:outermix}-\eqref{eq:innerrmix} are equivalent. 

\section{Inference}\label{sec:Inference}
Posterior inference can be performed through both a conditional and a marginal algorithm, derived by extending the algorithms by \citet{Argiento2022} to the enriched set-up. The conditional algorithm is described in Algorithm~\ref{algorithm}, while in Algorithm~\ref{alg:marg} we present the marginal one.

\begin{algorithm}[!htbp]
\caption{\label{algorithm}Conditional algorithm}
\hspace*{\algorithmicindent} \textbf{Input}: $\left(y_{ij}\right)_{ij}$ and parameter initialisation \\
\hspace*{\algorithmicindent} \textbf{Output}: posterior distribution of cluster allocation and other parameters\\
\begin{algorithmic}
	\For{$i$ in 1:$n$} 
	\State  Sample $c_i$ and $z_i$ from 
	\[
	\mathbb{P}[c_i = m, z_i = s \mid \text{rest}] \propto
	\Gamma_m \,\Delta_{ms}\, \prod_{j=1}^d f\left(y_{ij}\mid \bm{p}^\star_m, \bm{r}^\star_{ms}, \bm{\theta}^\star_{ms}\right)
	\]
	\EndFor
	\State Compute $K$, the number of allocated components at the outer level
	\State Relabel the outer level clusters so that the first $K$ components of the mixture are allocated
	\State Sample the latent variable $\bar u$ from $\text{Gamma}\left(n,\,\sum_{m=1}^M\Gamma_m\right)$
	\State Set $M = K + x$, where 
	\[
	\qquad x \sim q_x \qquad q_x \propto \frac{(x+K)!}{x!}\psi_{\textnormal{out}}(\bar u)^x q_M(K+x) \qquad \text{for } x=0,1,\ldots
	\]
	\State Sample the unnormalised weights of the outer measure from
	\[
	\mathbb{P}[\Gamma_m \in d\omega \mid \text{rest}] \propto \omega^{n_m} e^{-\omega \bar u}h_{\textnormal{out}}(\omega) d\omega
	\qquad \text{for } m=1,\ldots,M
	\]
	where $n_m$ is the cardinality of outer level cluster $m$ and $n_m=0$ for $m>K$
	\For{$m$ in 1:$K$}
	\State Sample $\bm p^\star_m$ from the full conditional.
	\State Compute the number $K_m$ of allocated components at the inner level
	\State Relabel the inner level clusters so that the first $K_m$ components are allocated
	\State Sample the latent variable $u_m$ from $\text{Gamma}\left(n_m,\,\sum_{s=1}^{S_m}\Delta_{ms}\right)$
	\State Set $S_m = K_m + x$ where 
	\[
	x \sim q_x \qquad q_x \propto \frac{(x+K_m )!}{x!}\psi_{\textnormal{in}}(u_m)^x q_S(K_m +x) \qquad \text{for } x=0,1,\ldots
	\]
	\State Sample the unnormalised weights of the $m$-th inner mixture from 
	\[
	\mathbb{P}[\Delta_{ms}\in dq\mid \text{rest}] \propto q^{n_{ms}} e^{-\omega u_m} h_{\textnormal{in}}(q) dq
	\qquad \text{for } s=1,\ldots,S_m
	\]
	$\quad\enskip$where $n_{ms}$ is the cardinality of inner level cluster $s$ and $n_{ms}=0$ for $s>K_m$
	\For{$s$ in 1:$K_m$}
	\State Sample $(\bm r^\star_{ms},\bm \theta^\star_{ms})$ from the full conditional
	\EndFor
	\For{$s$ in $(K_m+1)$:$S_m$}
	\State Sample $\bm r^\star_{ms}$ from the prior
	\State Sample $\theta^\star_{ms}$ from the prior
	\EndFor
	\EndFor
	\For{$m$ in $(K+1)$:$M$}
	\State Sample $\bm p^\star_m$ and $S_m$ from the prior
	\For{$s$ in 1:$S_m$}
	\State Sample $ \Delta_{ms}$ from the prior
	\State Sample $\bm r^\star_{ms}$ from the prior
	\State Sample $\theta^\star_{ms}$ from the prior
	\EndFor
	\EndFor
\end{algorithmic}
\end{algorithm}

\begin{algorithm}[h!] 
\caption{\label{alg:marg}Marginal algorithm }
\hspace*{\algorithmicindent} \textbf{Input}: $\left(y_{ij}\right)_{ij}$ and parameter initialisation \\
\hspace*{\algorithmicindent} \textbf{Output}: posterior distribution of cluster allocation and other posterior summaries\\
\begin{algorithmic}
	\For{$i$ in 1:$n$}
	\State Sample $c_i$ 
	\small
	\[
	\begin{aligned}
		\mathbb{P}&[c_i = m \mid \bm{c}^{-(i)}, \bm{z}^{-(i)}, \bar U, U_1, \ldots, U_K] \\
		&\propto\begin{cases}
			\begin{aligned}
				\left(n_m^{-(i)} + \gamma_M\right)\prod_{j=1}^d \frac{\mathcal{M}_\mathrm{Bern}\left(y^*_{j\mathcal{C}_m^{+(i)}}\right) }{\mathcal{M}_\mathrm{Bern}\left(y^*_{j\mathcal{C}_m^{-(i)}}\right)}\left(\frac{n_{ms}^{-(i)} + \gamma_S}{L^{-(i)}_m}
				\prod_{j=1}^d \frac{\mathcal{M}_\mathrm{NB}\left(y^*_{j\mathcal{C}_{ms}^{+(i)}} \right) 
				}{\mathcal{M}_\mathrm{NB}\left(y^*_{j\mathcal{C}_{ms}^{-(i)}} \right)}\right. \\
				\left. + \enskip \frac{L^{-(i)}_m - n^{-(i)}_m - \gamma_S}{L^{-(i)}_m} \prod_{j=1}^d \mathcal{M}_\mathrm{NB}\left(y_{ij}\right)\right)
			\end{aligned}& \text{if } m = m_\textnormal{old}\\
			\begin{aligned}
				\frac{\Lambda_M + (K^{-(i)} + 1)(\bar{u} + 1)^{\gamma_M}}{\Lambda_M + K^{-(i)}\left(\bar{u} + 1\right)^{\gamma_M}}\frac{\Lambda_M \, \gamma_M}{(\bar{u}+1)^{\gamma_M}}
				\prod_{j=1}^d\mathcal{M}_\mathrm{Bern}\left(y_{ij}\right)\mathcal{M}_\mathrm{NB}\left(y_{ij}\right)\end{aligned} & \text{otherwise }\\
		\end{cases}
	\end{aligned}
	\]
	\normalsize
	\indent where $n_m^{-(i)}$ and $n_{ms}^{-(i)}$ are the cardinalities of outer and inner clusters after removing the $i$-th observation, $\mathcal{C}_m^{-(i)} = \mathcal{C}_m \setminus \{i\} $ and  $\mathcal{C}_m^{+(i)} = \mathcal{C}_m \cup  \{i\}$, and similarly for  $\mathcal{C}_{ms}^{+(i)}$, $\mathcal{C}_{ms}^{-(i)}$, $K^{(-i)}$, and $K_m^{-(i)}$. Here the subscript `\text{old}' denotes an existing (occupied) cluster and
	\[
	L^{-(i)}_m = \frac{\Lambda_S + \left(K^{-(i)}_m + 1\right)\left(u_{m} + 1\right)^{\gamma_S}}{\Lambda_S + K^{-(i)}_m\left(u_{m}+ 1\right)^{\gamma_S}} \frac{\Lambda_S \, \gamma_S}{(u_{m}+1)^{\gamma_S}} + n_m^{-(i)} + \gamma_S
	\]
	\State Sample $z_i$ 
	\[
	\begin{aligned}
		\mathbb{P}&[z_i = s \mid \bm{c}, \bm{z}^{-(i)}, U_m ] \\
		&\propto\begin{cases}
			\begin{aligned}
				(n_{ms}^{-(i)} + \gamma_S)
				\prod_{j=1}^d \frac{\mathcal{M}_\mathrm{NB}\left(y^*_{j\mathcal{C}_{ms}^{+(i)}} \right) 
				}{\mathcal{M}_\mathrm{NB}\left(y^*_{j\mathcal{C}_{ms}^{-(i)}} \right)}
			\end{aligned}& \text{if } s = s_\textnormal{old}\\
			\begin{aligned}
				\frac{\Lambda_S + \left(K^{-(i)}_m + 1\right)\left(u_{m} + 1\right)^{\gamma_S}}{\Lambda_S + K^{-(i)}_m\left(u_{m}+ 1\right)^{\gamma_S}} \frac{\Lambda_S \, \gamma_S}{(u_{m}+1)^{\gamma_S}} \prod_{j=1}^d\mathcal{M}_\mathrm{NB}\left(y_{ij}\right)\end{aligned} & \text{otherwise }\\
		\end{cases}
	\end{aligned}
	\]
	\indent Note that when a subject $i$ is assigned to a new outer cluster, then the full conditional distribution of $z_i$ is degenerate and a new auxiliary variable $U_m$ has to be sampled before moving to the next subject $i+1$. 
	\EndFor
	\State Sample the latent variables $\bar{U}$ and $U_1,\ldots,U_K$ from their full conditional:
	\small
	\[
	\begin{aligned}
		\mathbb{P} \left[\bar{U} \in  \text{d}\bar{u}\mid \text{rest}\right] & \propto 
		\left(\frac{\Lambda_M}{\left(\bar{u} + 1\right)^{\gamma_M}} + K\right)
		\exp\left\{\frac{\Lambda_M}{\left(\bar{u} + 1\right)^{\gamma_M}}\right\}
		\frac{\bar{u}^{n-1}}{\left(\bar{u} + 1\right)^{n + K\gamma_M}},\quad \bar{u}>0\\
		\mathbb{P} \left[U_{m} \in  \text{d} u_{m}\mid \text{rest}\right] & 
		\propto 
		\left(\frac{\Lambda_S}{(u_{m} + 1)^{\gamma_S}} + K_m\right)
		\exp\left\{\frac{\Lambda_s}{\left(u_{m} + 1\right)^{\gamma_S}}\right\}
		\frac{\left(u_{m}\right)^{n_m-1}}{\left(u_{m} + 1\right)^{n_m + K_m\gamma_S}}, u_{m}>0
	\end{aligned}
	\]
	\normalsize
\end{algorithmic}
\end{algorithm}

The conditional algorithm is very flexible and allows for different prior distributions on the weights of the two mixtures as well as on $M$ and $S_m$ \citep[see][for details]{Argiento2022}. In Algorithm~\ref{alg:marg}, we use the notation $q_M$ and $q_S$ to denote the prior on $M$ and $S_m$, respectively, and we set them both equal to a shifted Poisson for the application in Section~\ref{sec:Application}. Furthermore, $h_{\textnormal{out}}$ and $h_{\textnormal{in}}$ denote the prior distribution on the unnormalised weights (in our case Gamma distributions) of the outer and inner mixture, respectively, $\psi_{\textnormal{out}}(u)$ and $\psi_{\textnormal{in}}(u)$ denote the corresponding Laplace transforms of $h_{\textnormal{out}}$ and $h_{\textnormal{in}}$ (in our case $\psi_{\textnormal{out}}(u) = (u + 1)^{-\gamma_M}$ and $\psi_{\textnormal{in}}(u) = (u + 1)^{-\gamma_S}$). 

To implement the marginal algorithm, we need to  derive the marginal likelihood of the data, conditionally on  cluster membership. The likelihood in Eq.~\eqref{eq:model_general} can be written as:
\begin{equation}\label{eq:1+2}
\begin{aligned}
	\prod_{i=1}^n \prod\limits_{j=1}^d 
	\Bigg\{\left(1-p_{ij}\right)^{1 - \widetilde{y}_{ij}} \, p_{ij}^{\widetilde{y}_{ij}}\,
	\bigg\{ \frac{\left(y_{ij} + r_{ij} - 2\right)!}{\left(r_{ij}-1\right)!\left(y_{ij}-1\right)!}
	\theta_{ij} ^{y_{ij}-1} \left(1-\theta_{ij}\right)^{r_{ij}} \bigg\}^{\widetilde{y}_{ij}}\Bigg\}
\end{aligned}
\end{equation}
Recall that $c_i$ and $z_i$ denote the labels of the clusters to which the $i$-th subject belongs to in the outer and the inner clustering, respectively. The marginal likelihood of the data conditionally on the cluster allocation is obtained marginalising with respect to the prior distributions defined in \eqref{eq:outermix} and \eqref{eq:innerrmix}. For a vector of counts $y$, we obtain: 

\small
\[
\begin{aligned}
&\mathcal{M}\left(y\mid \bm{c}, \bm{z}\right) =
\prod_{j=1}^{d}
\left\{\prod_{m=1}^K
\left\{\mathcal{M}_\mathrm{Bern}\left(y^*_{j \mathcal{C}_m}\right)\prod_{s=1}^{K_m} \mathcal{M}_\mathrm{NB}\left(y^*_{j \mathcal{C}_{ms}}\right)\right\}\right\} \\
&\mathcal{M}_\mathrm{Bern}\left(y\right) =\frac{B\left(\alpha + n^{1}, \beta + n^{0}\right)}{B(\alpha,\beta)}\\
&\mathcal{M}_\mathrm{NB}\left(y\right) =
\sum_{r=1}^{+\infty}
\left\{
\frac{B\left(\eta+ \sum_{i} \left(y_{i} - 1\right)\widetilde{y}_i,\lambda + r \, \sum_{i} \widetilde{y}_i\right)}{B\left(\eta,\lambda\right)} \prod\limits_{i} \left(\frac{\left(y_{i} + r - 2\right)!}{\left(r-1\right)!\left(y_{i}-1\right)!}\right)^{\widetilde{y}_i} \left(1-\zeta\right)^{r-1}\zeta \right\}
\end{aligned}
\]
\normalsize
where $\mathcal{C}_m = \{i: c_i=m\}$, $\mathcal{C}_{ms} = \{i: c_i=m,z_i=s\}$, $y^*_{j\mathcal{C}_m}$ is the vector of observations $y_{ij}$ such that $ c_i = m  $, for $j=1,\ldots,d$. Similarly, $y^*_{j \mathcal{C}_{ms}}$ is the vector of observations $y_{ij}$ such that $ c_i = m   $ and $ z_i=s$. Moreover, $B\left(\cdot,\cdot\right)$ denotes the Beta function, $n^{1} = \sum_{i} \widetilde{y}_{i}$, $n^{0} = \sum_{i} \left(1 - \widetilde{y}_{i}\right)$,  $\widetilde{y}_{i}$ is defined as in Eq.~\eqref{eq:ytilde} and the last two summations run over the elements of the vector $\widetilde{y}$. Here $K$ and $K_m$ are the numbers of clusters at the outer and inner level, respectively. Note that by cluster we mean an occupied component (i.e.\ a mixture component to which at least one observation has been assigned), with $K\leq M$ and $K_m\leq S_m,m=1,\ldots,M$. 

When implementing the marginal algorithm, after updating the latent variables $\bar U$ and $U_{m}$, we could add an extra step involving  a shuffle of the nested partition structure as suggested by ~\citet{wade2014improving} to improve mixing. More details and an empirical comparison of the two algorithms are provided in Section~S3 of Supplementary Material.


\section{Application to WhatsApp use during COVID-19}\label{sec:Application}

\subsection{Data description and preprocessing}

We apply our model to a dataset on WhatsApp use during COVID-19 \citep{clinicaltrials2021}. The data consist of a questionnaire filled out by participants living in India. Each subject answers the same $d=7$ questions for $T=7$ consecutive days on the number of 
($j=1$) COVID-19 messages forwarded,
($j=2$) WhatsApp groups to which COVID-19 messages were forwarded,
($j=3$) people to whom COVID-19 messages were forwarded,
($j=4$) unique forwarded messages received in personal chats,
($j=5$) people from whom forwarded messages were received,
($j=6$) personal chats that discussed COVID-19,
($j=7$) WhatsApp groups that mentioned COVID-19.
Table~S1 in Supplementary Material
provides the list of the questions, as well as a brief description. In what follows, the first replicate ($t=1$) corresponds to Sunday for all subjects, $t=2$ to Monday, up to $T=7$ corresponding to Saturday. The questionnaire responses were collected in June and July 2021, during India's infection wave of the Delta variant of the SARS-CoV-2 virus that causes coronavirus disease 2019 (COVID-19).

From the initial 1156 respondents, we remove two subjects for which no answers are available, resulting in a final sample size of $n=1154$. Moreover, 19\% of the observations are missing. We also treat counts higher than 400, which are very rare (7 observations out of 56 546), as missing data as they are very far from the range of the majority of the data. We handle missing data using a two-step procedure. Firstly, whenever possible, we recover missing zeros using deterministic imputation based on respondent's answers to other sections of the questionnaire. For instance, if the answer to the question ``did you send any message of this kind today?'' is ``no''  and there is a missing value for the question ``how many?'', we can reasonably assume that the answer to the latter question is zero. In this way, we can recover 0.5\% of the missing observations. Secondly, the remaining missing values are imputed using random forest imputation \citep[as implemented in the \texttt{R} package \texttt{mice},][]{mice}. In 
Section~S2 of Supplementary Material,
we provide more details on the data imputation technique and we present an empirical study to quantify the impact of data imputation on the results presented in the next section. Figure~S2 of Supplementary Material
displays the data after imputation. 

To account for the  fact that $T$ repeated observations are available for each subject and process, we need to slightly modify model~\eqref{eq:model_general}. We do so by assuming that the different time points are independent of each other, so that repeated observations can be straightforwardly included into the proposed model. Let $Y_{ijt}$ denote the count for the $i$-th subject and the $j$-th process at time $t$, $i=1,\ldots,n$, $j=1,\ldots,d$ and $t=1,\ldots,T $. We assume that $Y_{ijt}$ are conditionally independent, given the parameters of the model. Thus, the likelihood contribution of each subject $i$ is given by $\prod_{t=1}^T \prod_{j=1}^d f\left(y_{ijt} \mid \bm \psi^\star_{msj}\right)$. It must be highlighted that we are clustering individuals based on the pattern of all their observations, at each time point $t$ and for each process $j$. 

Finally we note that, thanks to the probabilistic structure of the hurdle model for zero-inflated data, $\bm p_i$ and the sampling distribution $g\left(\cdot \mid \bm \mu_i\right)$ reflect two distinct features of the respondents' behaviour: $\bm p_i$ represents the probability of engaging in some COVID-19 related WhatsApp activity, while $g\left(\cdot \mid \bm \mu_i\right)$ captures the behaviour of those subjects who have actually engaged in the activity.

\subsection{Results}

Posterior inference is performed through the conditional algorithm described in Algorithm~\ref{algorithm}.  We run the algorithm for 15 000 MCMC iterations, discarding the first 5000 as burn-in. 

\begin{figure}[tb!]
\centering
\begin{subfigure}[]{0.4\textwidth}
	\includegraphics[width=\textwidth]{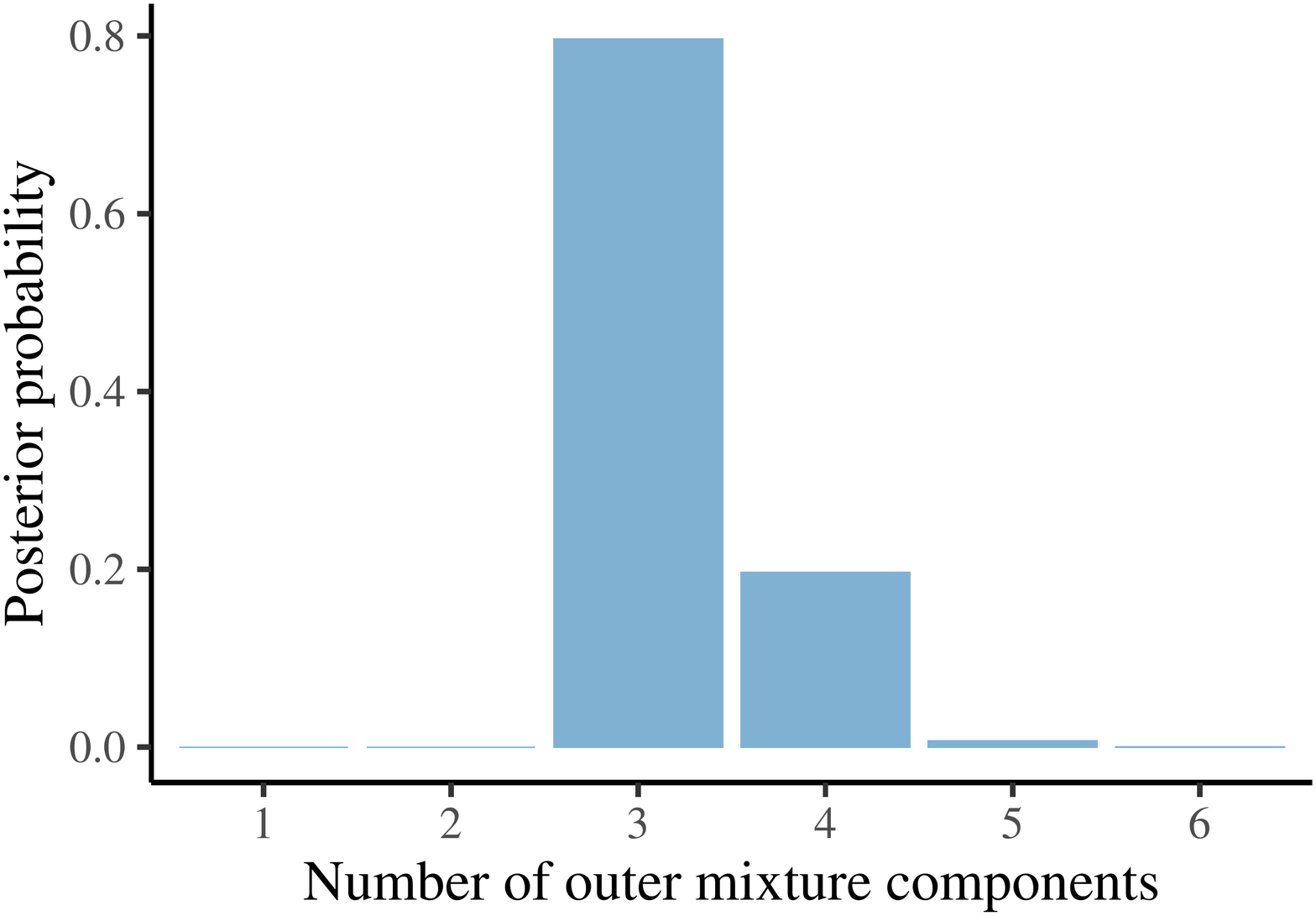}
\end{subfigure}
\hfill
\begin{subfigure}[]{0.4\textwidth}
	\includegraphics[width=\textwidth]{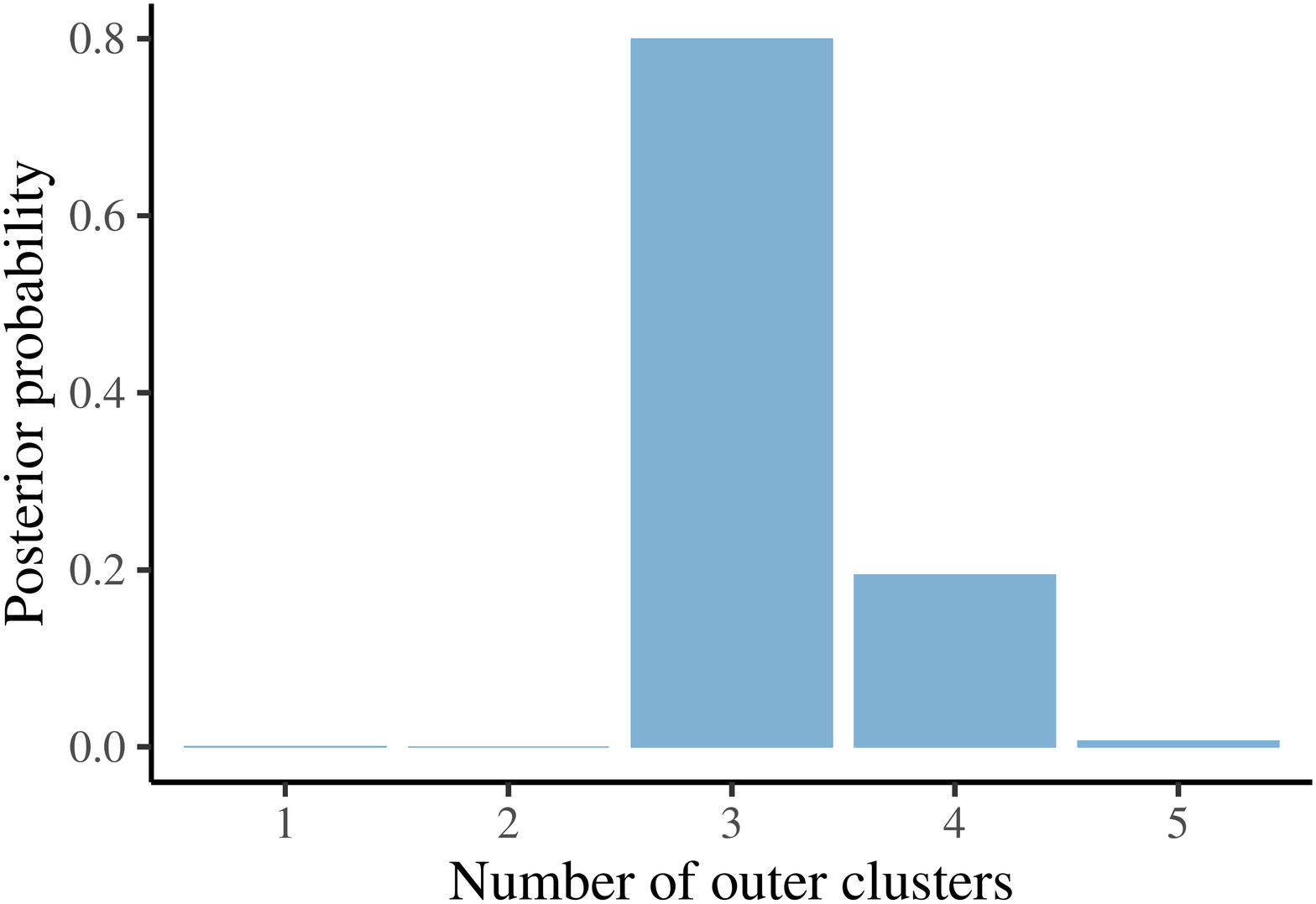}
\end{subfigure}
\caption{\label{fig:up}Posterior distribution of the number of outer mixture components $M$ (left panel) and clusters $K$, i.e., number of occupied components to which at least one observation is assigned (right panel).}
\end{figure}

Figure~\ref{fig:up} shows that, at the outer level, the posterior distributions of the number of both components and clusters present a mode at the value three.

\begin{figure}[tb!]
\centering

\begin{subfigure}[]{0.4\textwidth}
	\includegraphics[width=\textwidth]{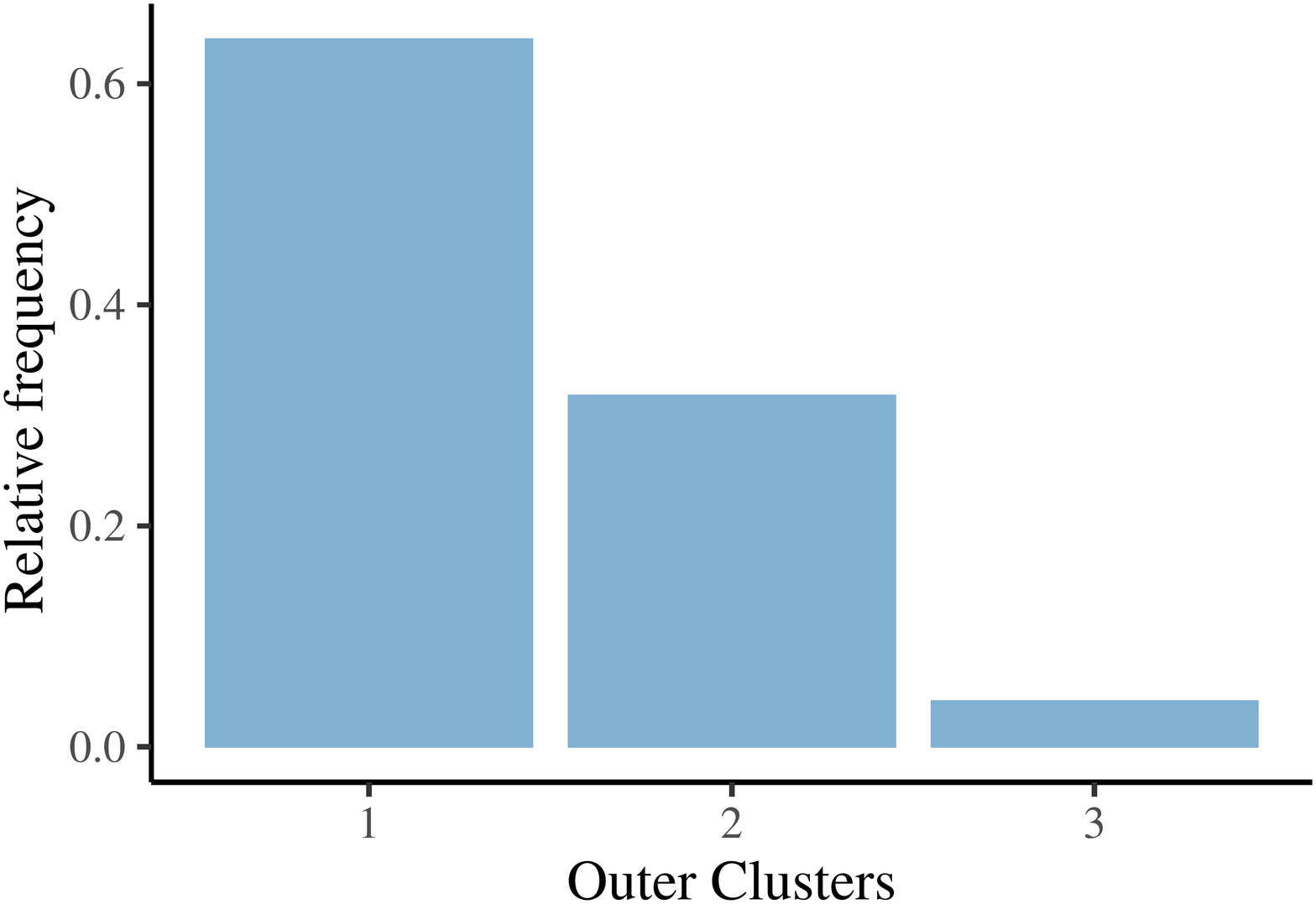}
	\subcaption{Outer clusters relative frequencies}
\end{subfigure}
\hfill
\begin{subfigure}[]{0.5\textwidth}
	\includegraphics[trim={6.5cm 5cm 6.5cm 6cm},clip,width=\textwidth]{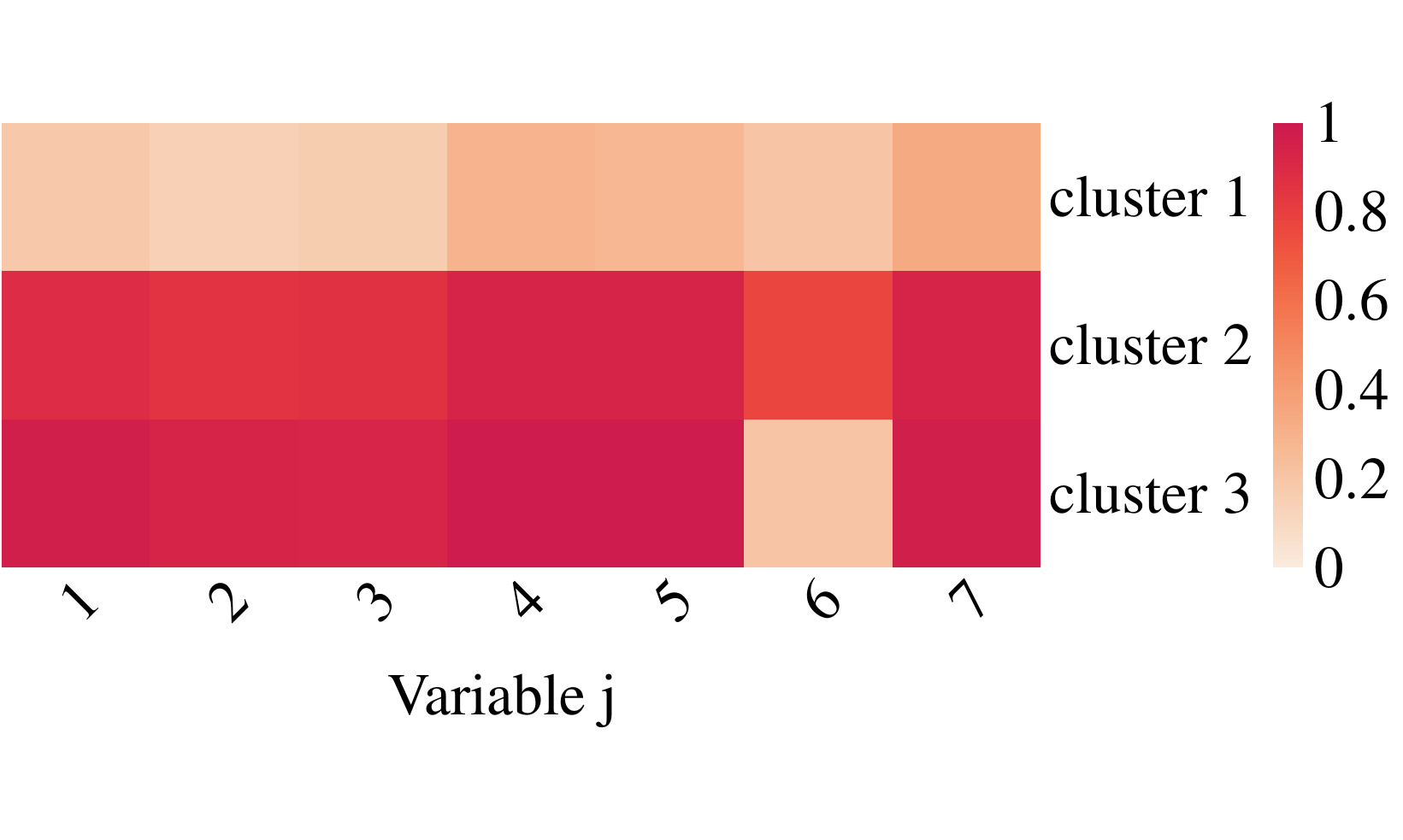}
	\subcaption{Outer level Bernoulli parameters $p^\star_{mj}$}
\end{subfigure}
\caption{\label{fig:up_binder}Relative frequency of the outer clusters (left panel) and the posterior means of the cluster-specific probabilities of a non-zero count $p^\star_{mj}$ (right panel) corresponding to the posterior estimate of the clustering allocation obtained by minimising Binder's loss function.}
\end{figure}

\begin{figure}[tb!]
\centering
\begin{subfigure}[]{0.32\textwidth}
	\includegraphics[trim={0cm 3cm 0cm 0cm},clip,width=\textwidth]{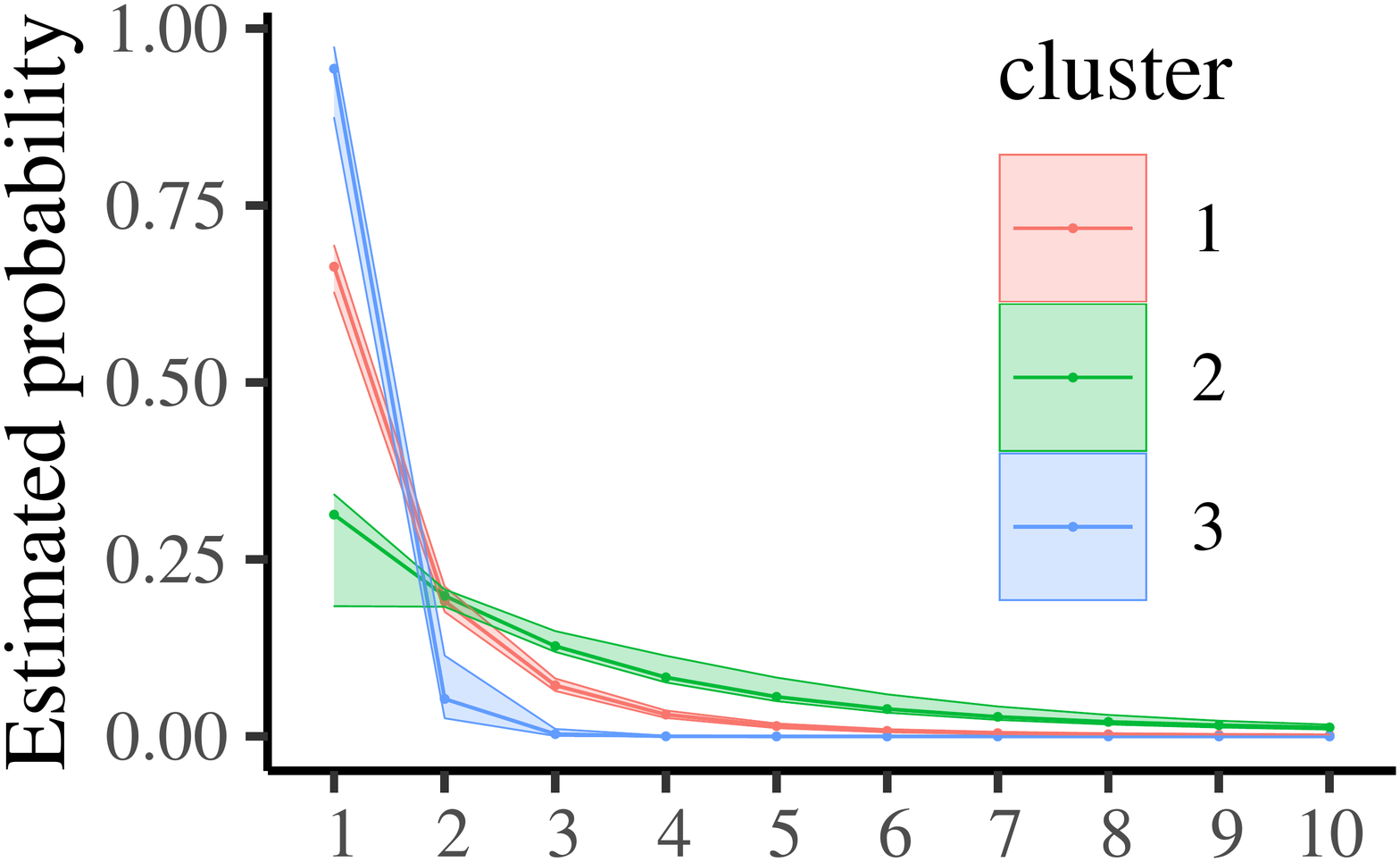}
	\subcaption{Question 1}
\end{subfigure}
\hfill
\begin{subfigure}[]{0.32\textwidth}
	\includegraphics[trim={0cm 3cm 0cm 0cm},clip,width=\textwidth]{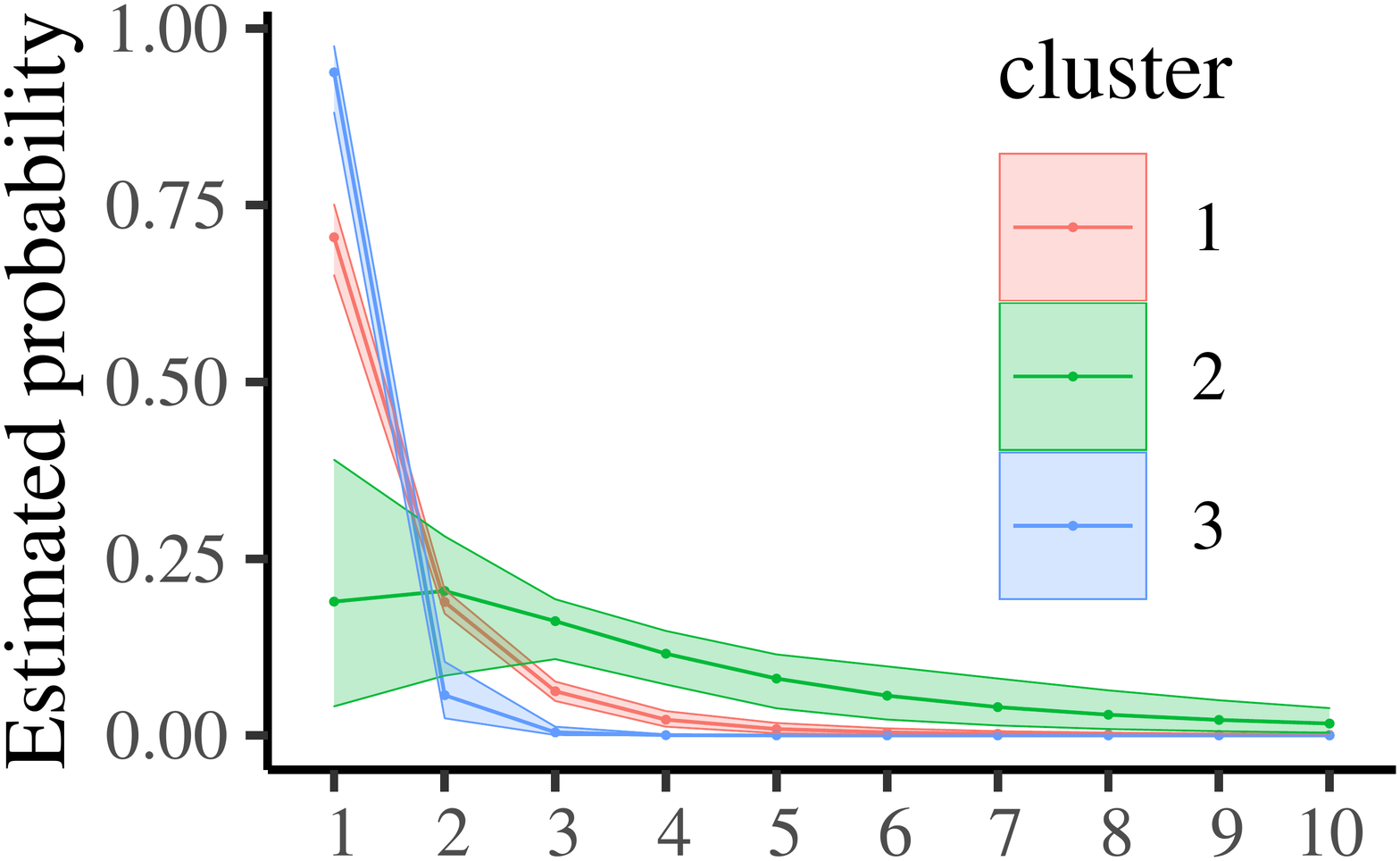}
	\subcaption{Question 2}
\end{subfigure}
\hfill
\begin{subfigure}[]{0.32\textwidth}
	\includegraphics[trim={0cm 3cm 0cm 0cm},clip,width=\textwidth]{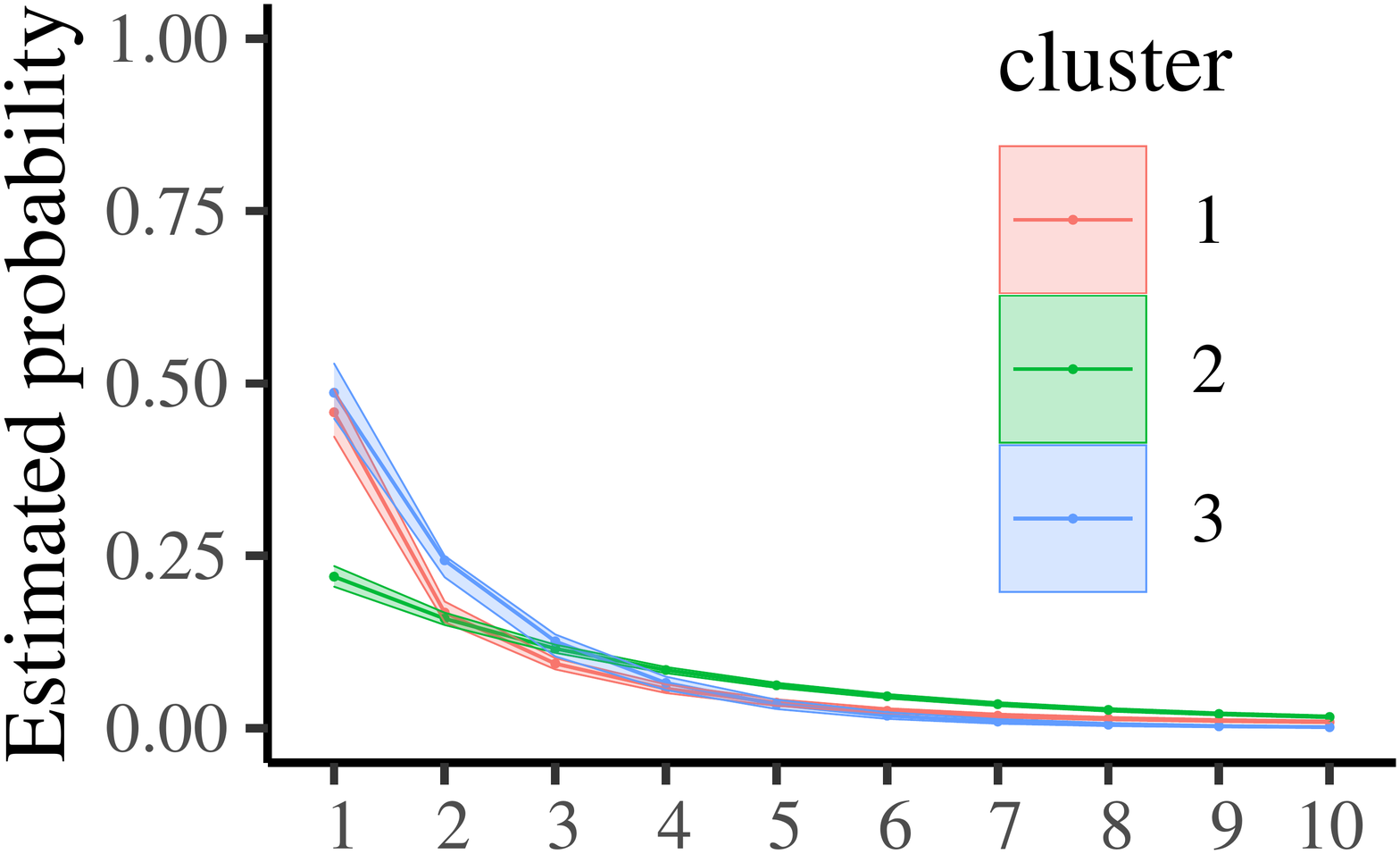}
	\subcaption{Question 3}
\end{subfigure}
\begin{subfigure}[]{0.245\textwidth}
	\includegraphics[trim={0cm 3cm 0cm 0cm},clip,width=\textwidth]{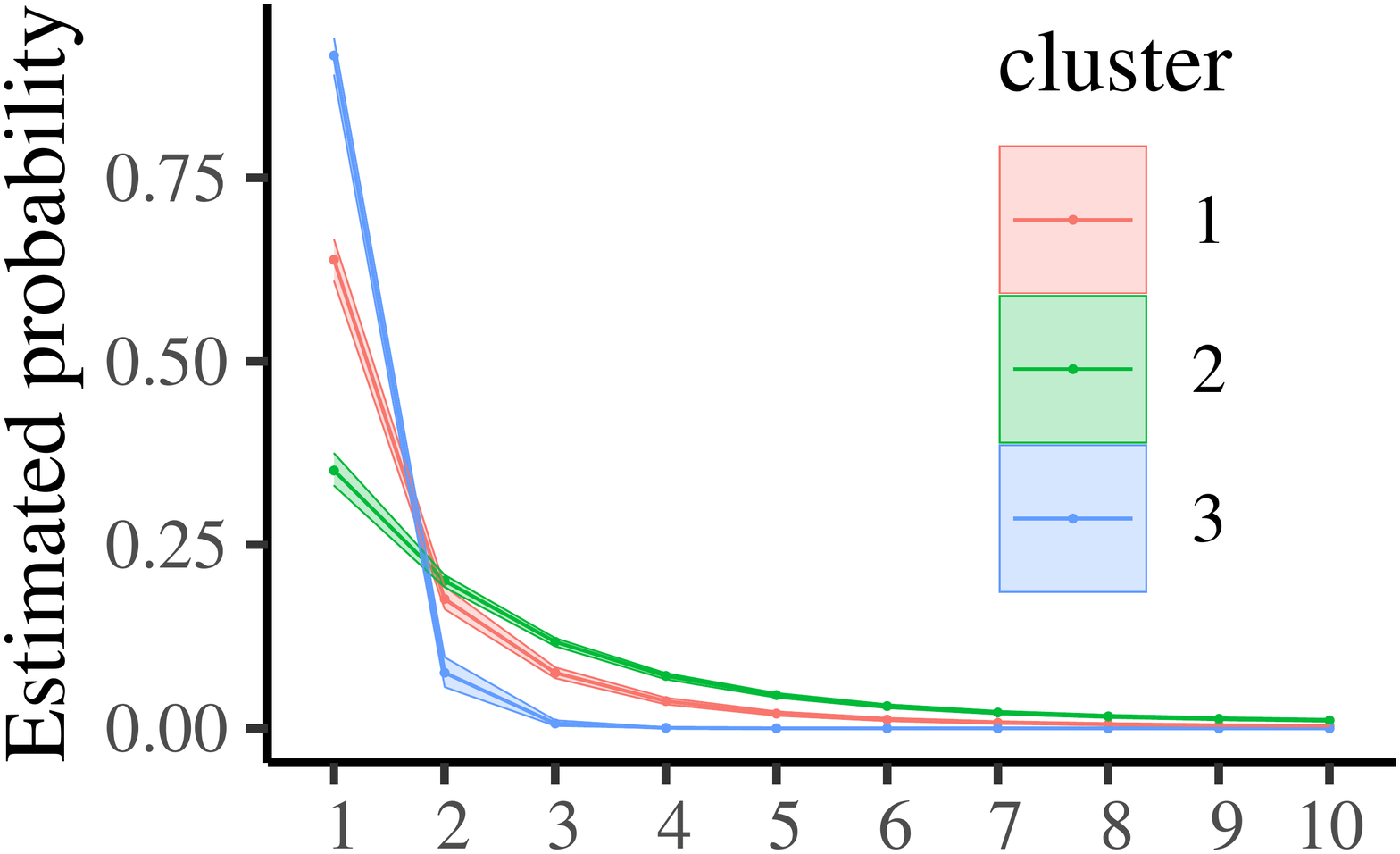}
	\subcaption{Question 4}
\end{subfigure}
\hfill
\begin{subfigure}[]{0.245\textwidth}
	\includegraphics[trim={0cm 3cm 0cm 0cm},clip,width=\textwidth]{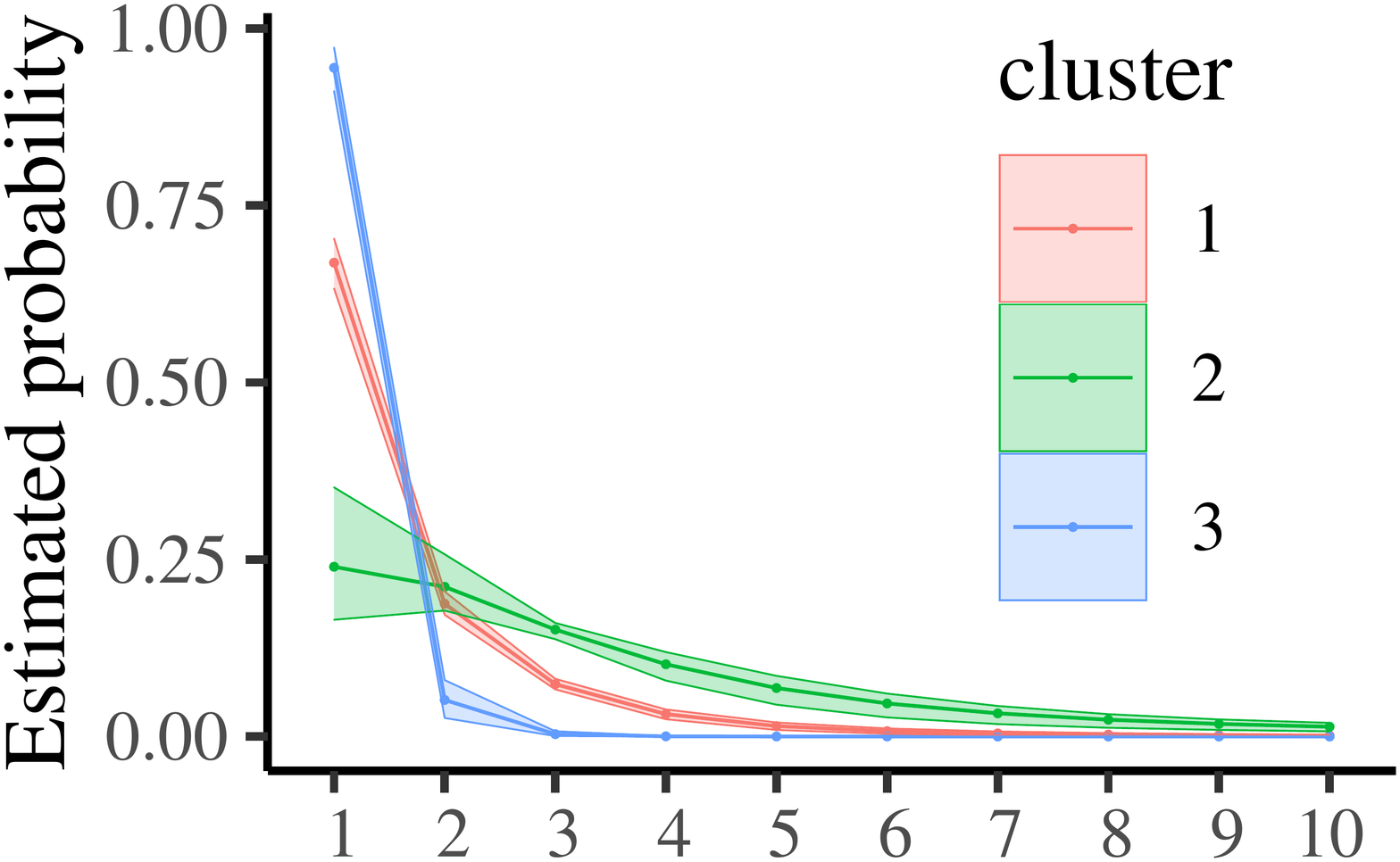}
	\subcaption{Question 5}
\end{subfigure}
\hfill
\begin{subfigure}[]{0.245\textwidth}
	\includegraphics[trim={0cm 3cm 0cm 0cm},clip,width=\textwidth]{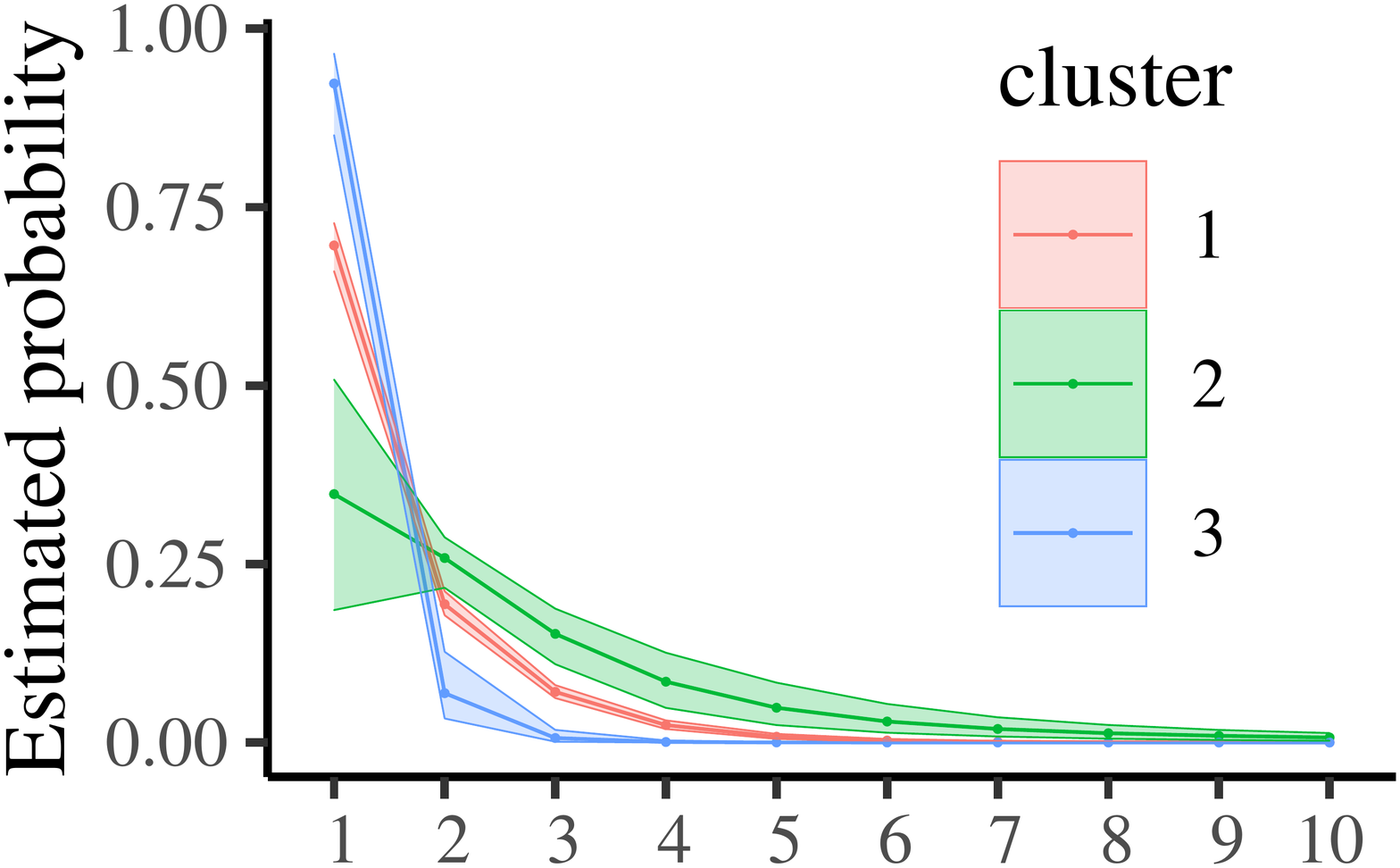}
	\subcaption{Question 6}
\end{subfigure}
\begin{subfigure}[]{0.245\textwidth}
	\includegraphics[trim={0cm 3cm 0cm 0cm},clip,width=\textwidth]{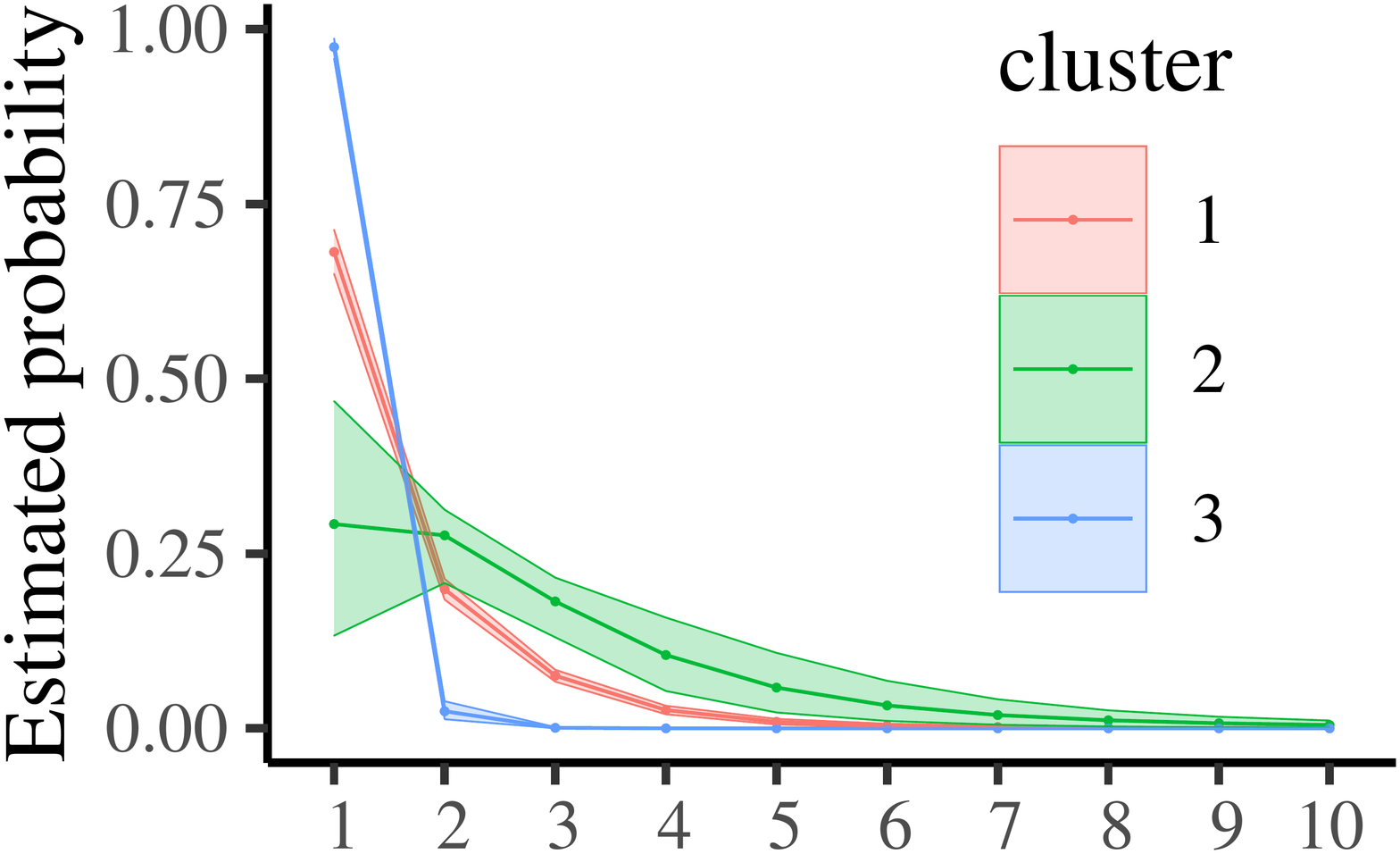}
	\subcaption{Question 7}
\end{subfigure}
\caption{\label{fig:pmf}
	Estimated pmfs for the seven questions within each outer cluster (conditionally on the counts being positive) corresponding to the posterior estimate of the clustering allocation obtained by minimising Binder's loss function. Shaded areas represent the 95\% credible intervals.
}
\end{figure}

As point estimate of the cluster allocation, we report the configuration that minimises the posterior expectation of Binder's loss function \citep{Binder1978} under equal misclassification costs, which is a common choice in the applied Bayesian nonparametrics literature \citep{Lau2007}. Briefly, this expectation of the loss measures the difference for all possible pairs of subjects between the posterior probability of co-clustering and the estimated cluster allocation. We refer to the resulting cluster allocation as the  Binder estimate.

The Binder estimate of the outer clustering contains three clusters, whose characteristics are summarised in Figures~\ref{fig:up_binder} and \ref{fig:pmf}. The largest cluster corresponds to WhatsApp users who on most days report a zero count for all $d=7$ questions. The individuals in the other two clusters use WhatsApp more frequently when it comes to forwarding COVID-19 messages ($j=1,2$), receiving forwarded messages ($j=3,4,5$) and having COVID-19 mentioned in their WhatsApp groups ($j=7$). The main feature distinguishing Cluster~2 from Cluster~3 in terms of probabilities $\bm p_i$ of non-zero counts is that on most days Cluster~2, unlike Cluster~3, discusses COVID-19 also in personal chats (question $j=6$).

\begin{figure}[!tb]
\centering
\begin{subfigure}[]{\textwidth}
	\includegraphics[trim={0cm 12cm 0cm 0cm}, clip,width=\textwidth]{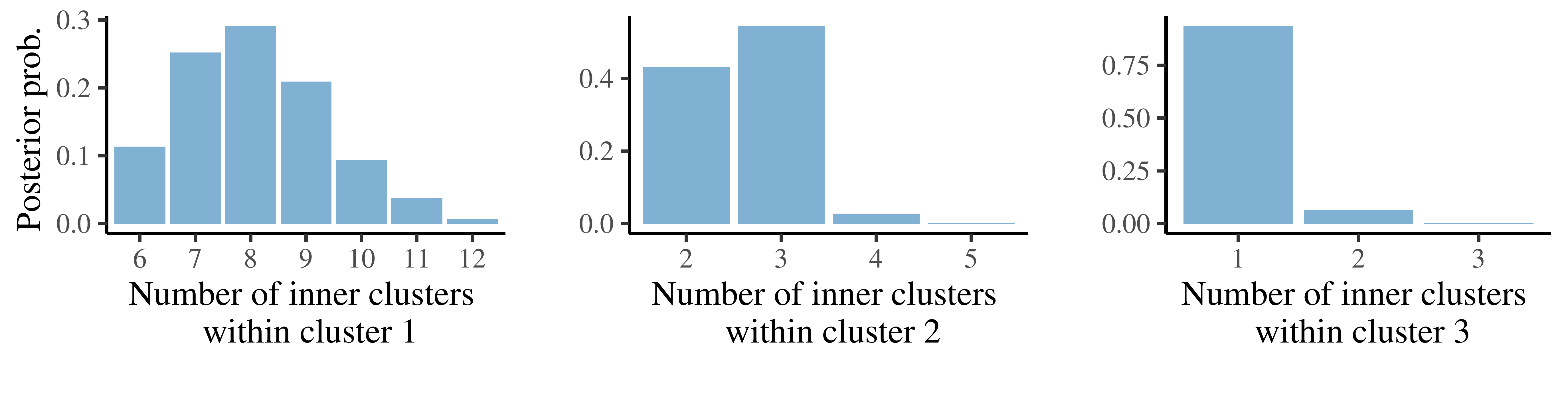}
\end{subfigure}
\begin{subfigure}[]{\textwidth}
	\includegraphics[trim={0cm 12cm 0cm 0cm}, clip,width=\textwidth]{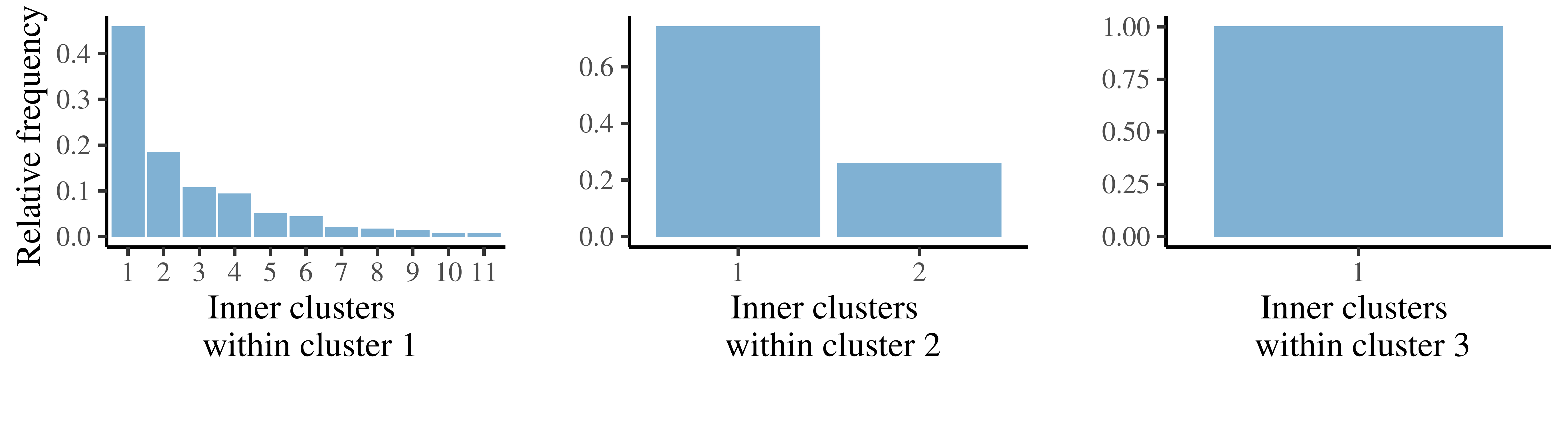}
\end{subfigure}
\begin{subfigure}[]{0.65\textwidth}
	\includegraphics[trim={0cm 7cm 0cm 0cm}, clip, width=\textwidth]{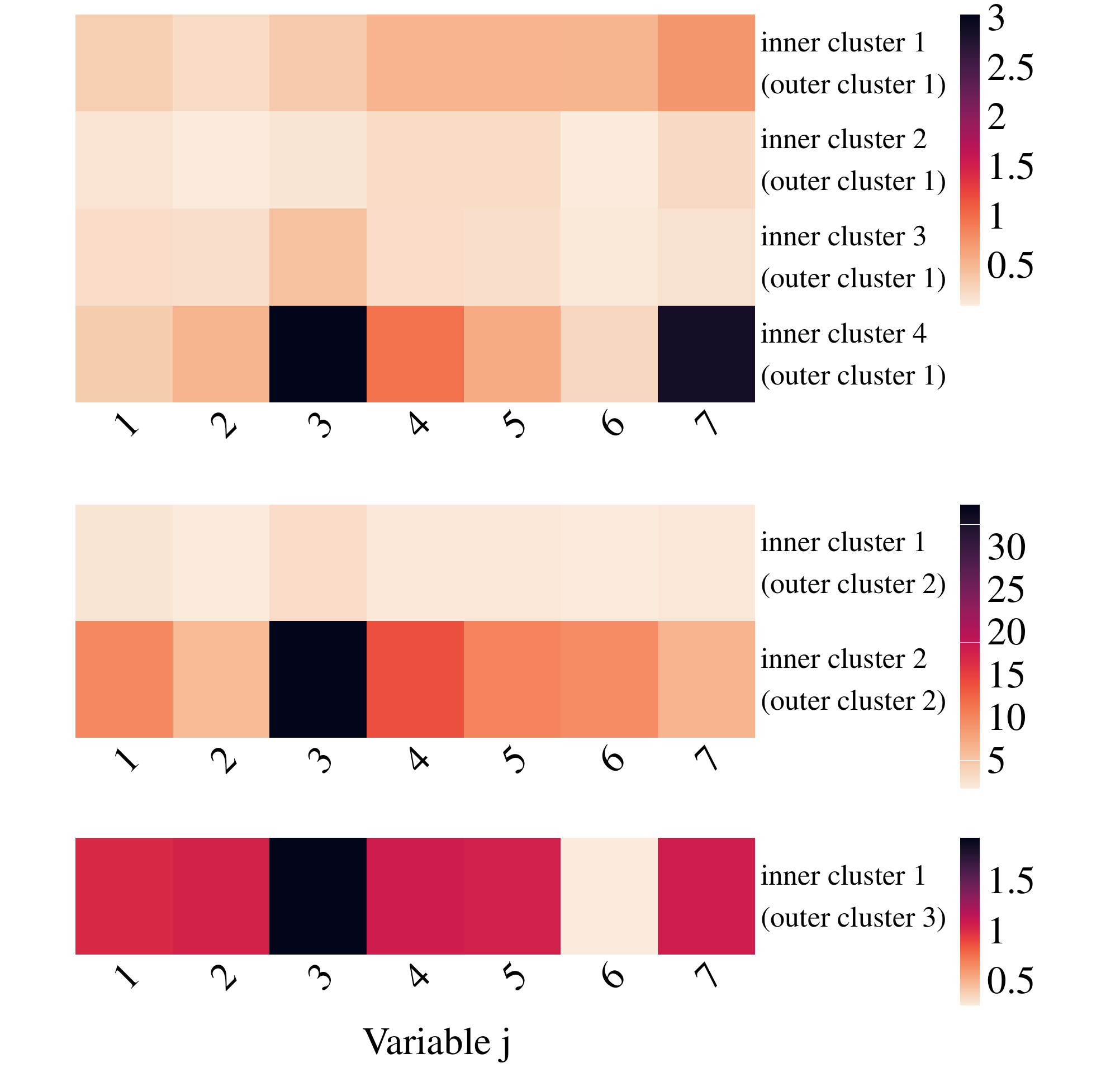}
\end{subfigure}
\caption{\label{fig:low}
	Posterior distribution of the number of inner clusters per outer cluster (top panel),
	relative frequency of the inner clusters corresponding to the Binder estimate of the inner cluster allocation (central panel), cluster-specific empirical means of the counts (bottom panel). For outer Cluster~1, the latter is only shown for the four largest inner clusters for visualisation purposes. Results are obtained conditionally on the Binder estimate of the outer clustering.
}
\end{figure}

\begin{figure}[!tb]
\centering
\begin{subfigure}[]{0.3\textwidth}
	\includegraphics[width=\textwidth]{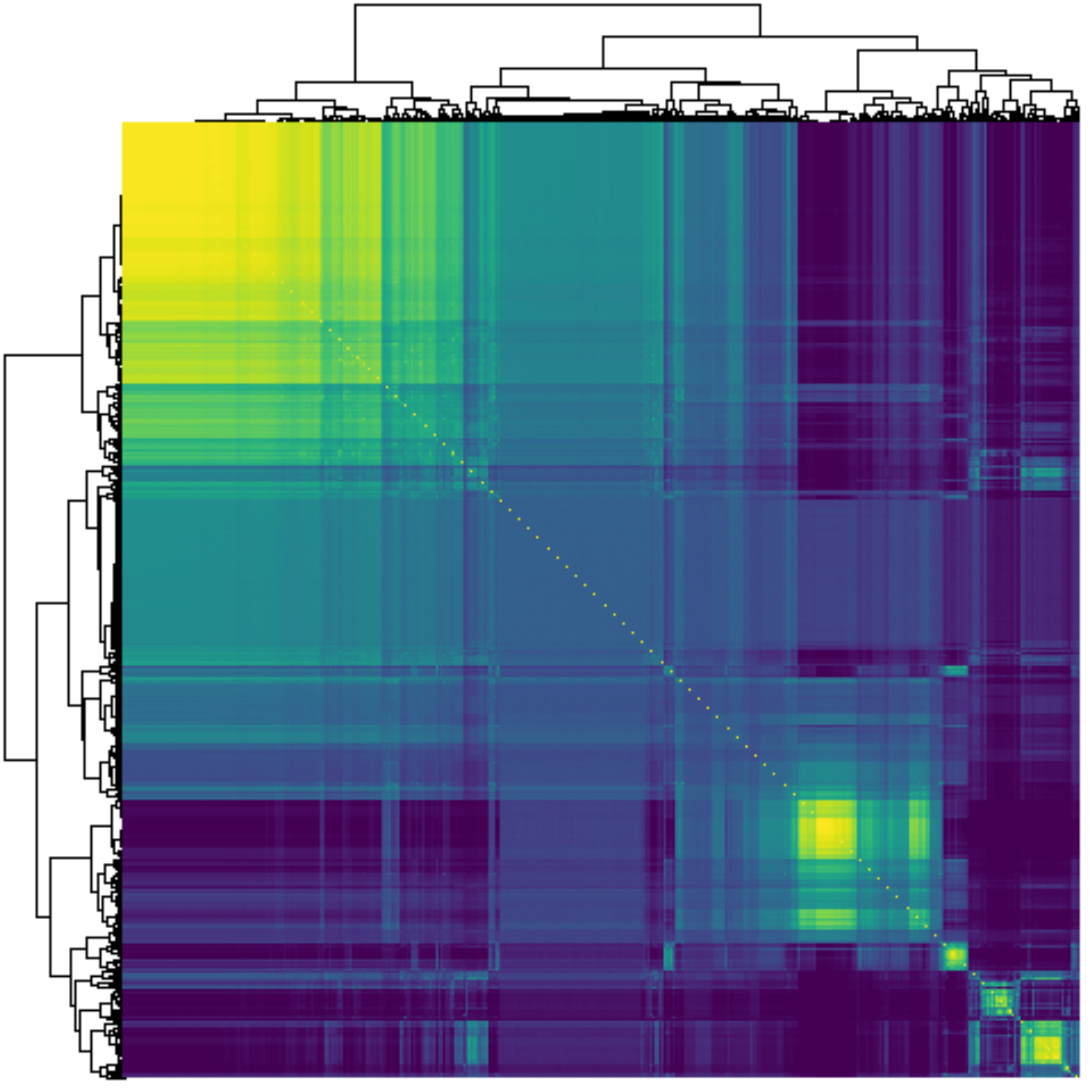}
	\subcaption{Outer cluster 1}
\end{subfigure}
\hfill
\begin{subfigure}[]{0.3\textwidth}
	\includegraphics[width=\textwidth]{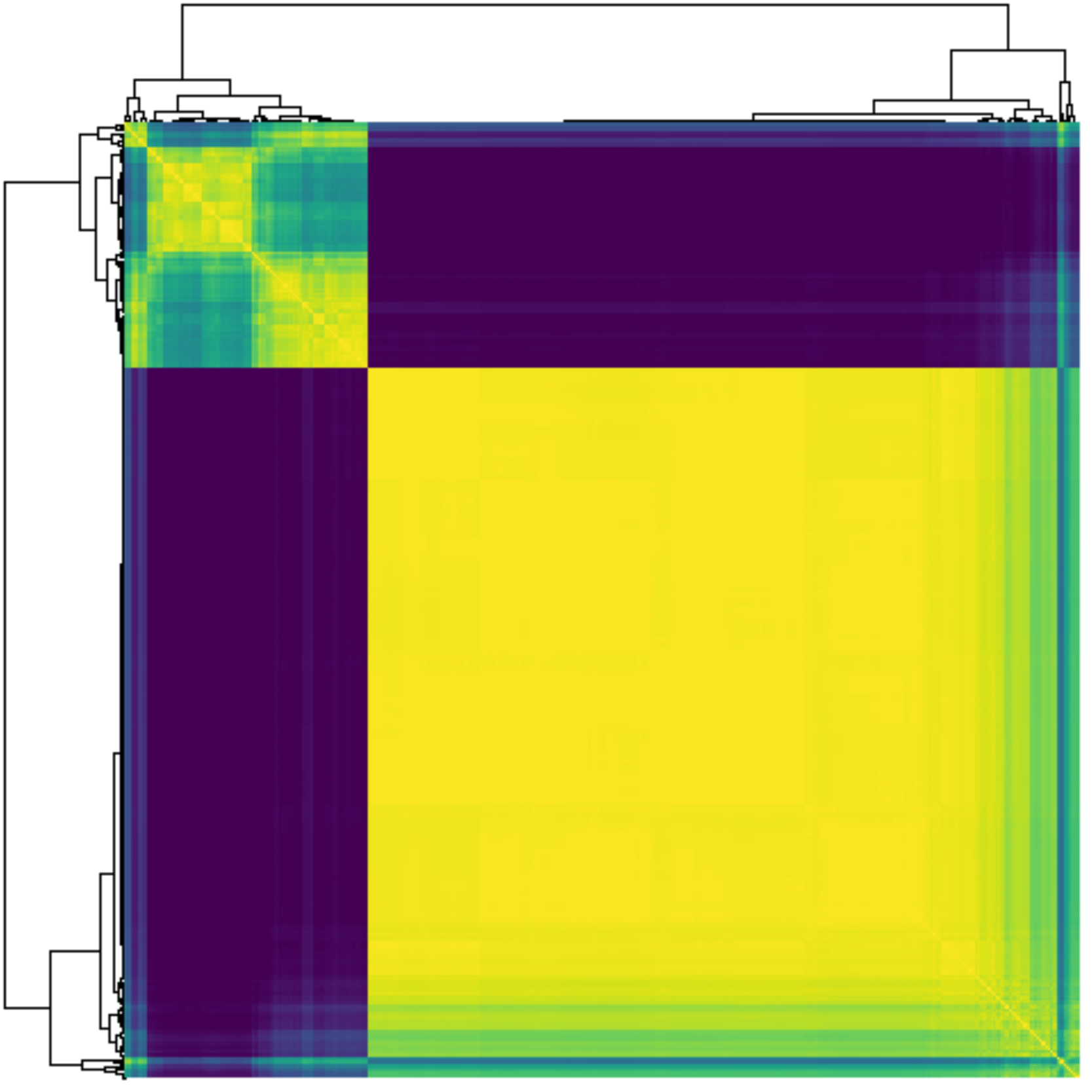}
	\subcaption{Outer cluster 2}
\end{subfigure}
\hfill
\begin{subfigure}[]{0.3\textwidth}
	\includegraphics[width=\textwidth]{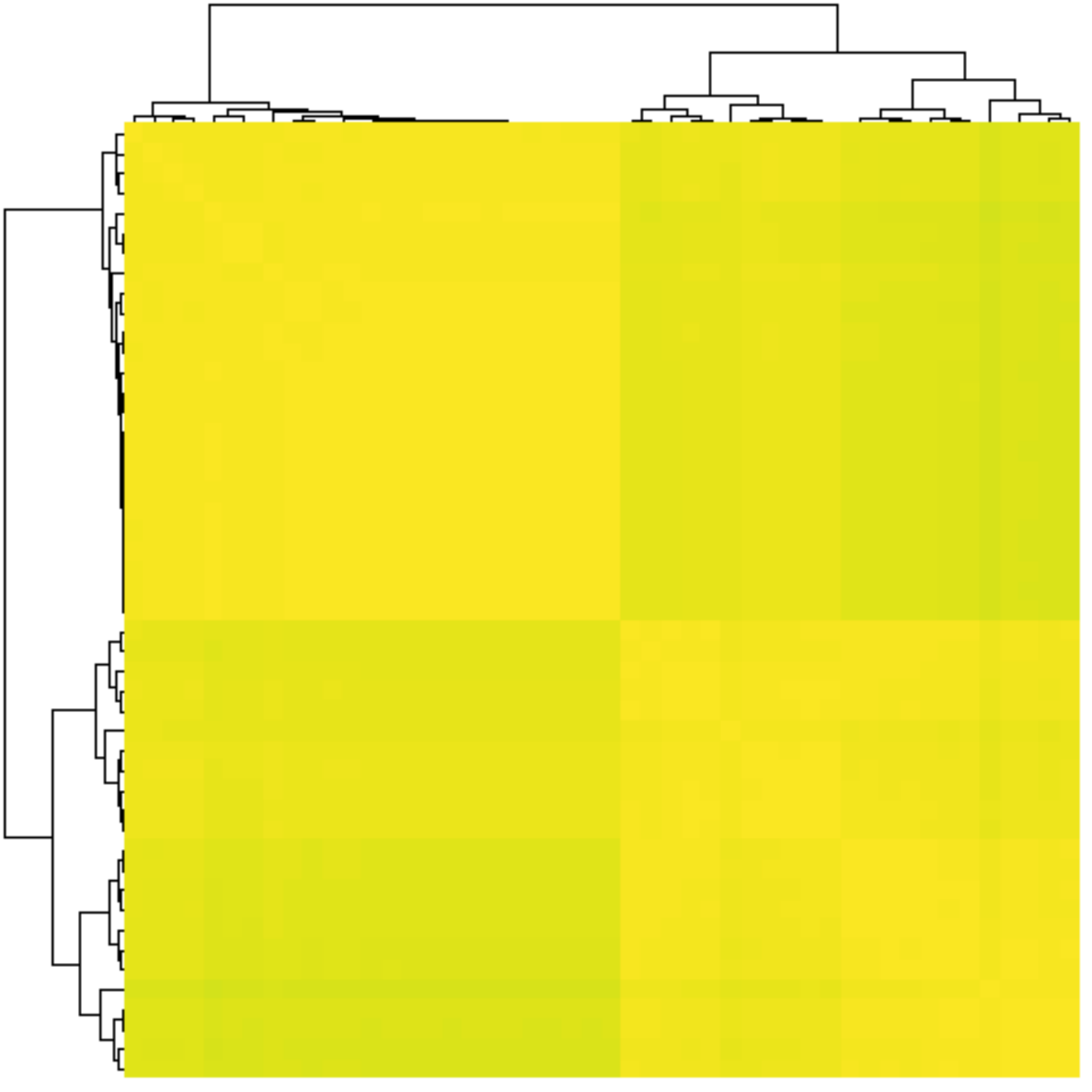}
	\subcaption{Outer cluster 3}
\end{subfigure}
\begin{subfigure}[]{0.075\textwidth}
	\includegraphics[width=\textwidth]{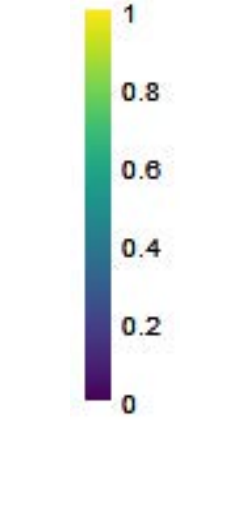}
\end{subfigure}
\caption{\label{fig:heat}
	Heatmaps of the posterior co-clustering probabilities for the inner clusters per outer cluster.
	Results are obtained conditionally on the Binder estimate of the outer cluster allocation.
	Observations are reordered based on the co-clustering probability profiles, through hierarchical clustering. 
}
\end{figure}

Figures~\ref{fig:low} and \ref{fig:heat} display the main characteristics of the inner clusters.
We are interested in the posterior distribution of the number of the inner clusters per outer cluster, as well as the inner clustering within each outer cluster. To this end, we run the MCMC algorithm fixing the outer cluster allocation to its Binder estimate, thus obtaining the conditional posterior distribution of the inner clustering. The results reveal substantial variability in the distribution of non-zero counts within outer Clusters~1 and 2 (see Figure~\ref{fig:low}, bottom panel).
The majority of counts in outer Cluster~1 are zero, leaving little variation in the counts for the inner clustering. As most individuals present zero counts (for most processes) at an inner cluster level, it becomes difficult to detect specific patterns as it is also evident from the fact that many co-clustering probabilities are in the range 0.3-0.6 (see Figure~\ref{fig:heat}). 
Notably, around a quarter of the individuals in outer Cluster~2, as captured by its inner Cluster~2, forward COVID-19 messages to many more people (question $j=3$) than subjects in inner Cluster~1 of outer Cluster~2. Figure~\ref{fig:pmf} also supports the fact that outer Cluster~2 engages with WhatsApp in a much more persistent manner than the other outer clusters. These results highlight that a sizeable minority of WhatsApp users
has a relatively large propensity to spread COVID-19 messages during a critical phase of the pandemic. This is in line with a similar survey in Singapore \citep{Tan2021} and findings on ``superspreaders'' on other social media.

\section{Conclusion}\label{sec:Concludes}
In this work, we propose a Bayesian  model for multiple zero-inflated count data, building on the well-established hurdle model and exploiting the flexibility of finite mixture models with random number of components. The main contribution of this work is the construction of an {\em enriched} finite mixture with random number of components, which allows for two level (nested) clustering of the subjects based on their pattern of counts across different processes. This structure enhances interpretability of the results and has the potential to better capture important features of the data. We design a conditional and a marginal MCMC sampling scheme to perform posterior inference. The proposed methodology has wide applicability, since excess-of-zeros count data arise in many fields. 
Our motivating application involves answers to a questionnaire on the use of WhatsApp in India during the COVID-19 pandemic. Our analysis identifies a two-level clustering of the subjects: the outer cluster allocation reflects daily probabilities of engaging in different WhatsApp activities, while the inner level informs on the number of messages conditionally on the fact that the subject is indeed receiving/sending messages on WhatsApp. 
Any two subjects are clustered together if they show a similar pattern across the multiple responses. 
We find three different well-distinguished respondent behaviours corresponding to the three outer clusters: (i) subjects with low probability of daily utilisation; (ii) subjects with high probability of sending/receiving all types of messages and (iii) subjects with high probability for all considered messages except for non-forwarded messages in personal chats. Interestingly, the inner level clustering and the outer cluster specific estimates of the sampling distribution $g$ highlight similarities between the outer Clusters~1 and 3, where subjects tend to send/receive fewer messages compared to outer Cluster~2. Moreover, we are able to identify those subjects with high propensity to
spread COVID-19 messages during the critical phase of the pandemic and for these subjects we do not find notable differences in terms of types of messages sent or received.
Our results are in line with existing literature on the topic. Future work involves the development of more complex clustering hierarchies and techniques able to identify processes that most inform the clustering structure.

\textbf{Funding:}{ This work was partially supported by the NUS Centre for Trusted Internet and Community [grant number CTIC-RP-20-09].}

\textbf{Acknowledgements:}{ We thank Dr.~Jean Liu and the Synergy Lab at Yale-NUS College for providing the data.}

\bibliographystyle{chicago}
\bibliography{references.bib}

\newpage
\begin{changemargin}{-1.5cm}{-1.5cm} 

\begin{center}
\Large{Supplementary material  for \\ ``Bayesian clustering of multiple zero-inflated outcomes''\\ by \\
	\normalsize Beatrice Franzolini, Andrea Cremaschi, Willem van den Boom and Maria De Iorio}
\end{center}

\normalsize

\section*{ S1 Questionnaire questions} \label{ap:q}
\begin{table}[h!] 
	\caption{\label{tab:response} Items in the questionnaire on WhatsApp activity.  Each question corresponds to a process in the model.}
	\begin{tabular}{c|c|c|l}
		&&Type of & \\
		Index $j$& Behaviour & Message&Question\\
		\hline\hline
		1&Sender&Forwarded&\small{How many different COVID-19 messages did you }\\
		&&&\small{ forward today?}\\
		2&Sender&Forwarded&\small{How many different WhatsApp groups did you }\\
		&&&\small{ forward COVID-19 messages to today?}\\
		3&Sender&Forwarded&\small{How many different people did you forward }\\
		&&&\small{ COVID-19 messages to today?}\\
		4&Recipient &Forwarded&\small{How many unique forwarded messages did you }\\
		&&&\small{ receive in your personal chats?}\\
		5&Recipient &Forwarded&\small{How many different people did you receive }\\
		&&&\small{ forwarded messages from today?}\\
		6&Participant &Personal comment&\small{How many of your personal chats discussed }\\
		&&&\small{ COVID-19 today?}\\
		7&Recipient &Both&\small{How many of your WhatsApp groups mentioned}\\
		&&&\small{ COVID-19 today?}
	\end{tabular}
\end{table}

\section*{ S2 Data imputation} \label{ap:dataimpute}
Missing data are imputed via multivariate imputation by chained equations as implemented in the \texttt{R} package \texttt{mice} \cite{mice}.  The method can be summarised as follows:  for each variable presenting missing entries a conditional distribution is estimated given all  other variables. To this end, random forests are employed. Missing values are then imputed from such distribution. Imputation is  iterated five times conditionally on previous values. More details on the procedure can be found in \cite{mice}.

In Table~\ref{tab:imputation} we compare results presented in the main manuscript with those obtained from other four  imputed datasets. The table contains the following posterior inference summaries: maximum a posteriori (MAP)  estimates of  the number of outer clusters (MAP-est $K$) and of the total number of inner clusters (MAP-est $\sum_m K_m$); Hellinger distance (H-dist) and Jensen-Shannon Divergence (JSD) between the posterior distribution of $K$ and $\sum_m K_m$ from the analysis presented in the main manuscript and the distribution obtained from each replicate; two adjusted rand indexes between the outer  partition in the main manuscript and the outer partition obtained from each of the replication dataset. We estimate the partition by (i) minimising the Binder loss estimate of the outer clustering (Adj Rand index Binder) (ii) applying the clustering method of Medvedovic (Adj Rand index Medvedovic), as implemented in the \texttt{R} package \texttt{mcclust}.

The Hellinger distance and the Jensen-Shannon divergence take values between 0 and 1, with higher values corresponding to more dissimilarities between the two distributions. The adjusted rand index varies between 0 and 1, with higher values indicating higher similarities between clustering allocations.

\begin{table}[h] 
	\caption{\label{tab:imputation} Impact of data imputation.}
	\begin{tabular}{l|c|c|c|c|c}
		&Main Analysis& Replica1 & Replica 2 & Replica 3 & Replica 4\\
		\hline
		MAP-est $K$ & 3& 3 & 3 & 3 & 3\\
		MAP-est $\sum_m K_m$&13&13&10&11&13\\
		\hline
		H-dist $K$ & - & 0.309 & 0.425 & 0.269 & 0.307\\
		JSD $K$& - & 0.101&0.189&0.081&0.098\\
		H-dist $\sum_m K_m$ & - & 0.299&0.453&0.323&0.285 \\
		JSD $\sum_m K_m$& - & 0.113 & 0.249&0.128&0.106 \\
		\hline
		Adj Rand index Binder & - & 0.732& 0.679&0.720&0.781\\
		Adj Rand index Medv. & - & 0.749&0.722 & 0.749&0.807\\
	\end{tabular}
\end{table}

\section*{ S3 Marginal algorithm} \label{ap:marg}
We compare the performance of the two sampling schemes presented in Section~3 
(i.e., the conditional algorithm and the marginal algorithm) on the dataset on WhatsApp use during COVID-19. 

We perform posterior inference employing both MCMC schemes  and  compare them in terms of effective sample size per iteration (ESS/\#iter) and integrated autocorrelation time (IAT) computed with the \texttt{R} packages \texttt{coda} and \texttt{LaplacesDemon}, respectively. ESS/\#iter and IAT are computed for the number of outer clusters, the number of inner clusters, and the log-likelihood $\log \mathcal{M}(y\mid \bm{c}, \bm{z})$. Results for the conditional algorithm are based on 10000 iterations. Results for the marginal algorithm are obtained with 1000 iterations.

Table~\ref{tab:condvsmarg} highlights that the marginal algorithm, as expected, leads to higher effective sample sizes per iteration as a consequence of the smaller number of parameters to sample and the lower dependence across them. This  ultimately allows for a better mixing of the chain.

\begin{table}[h] 
	\caption{\label{tab:condvsmarg} Comparison between conditional and marginal algorithm.}
	\resizebox{\columnwidth}{!}{%
		\begin{tabular}{l|c|c|c|c|c|c|}
			& \multicolumn{2}{c|}{Log-likelihood}& 
			\multicolumn{2}{c|}{Num. of outer clusters} & 
			\multicolumn{2}{c|}{Num. of inner clusters} \\
			&$\quad$Cond.$\quad$&$\quad$ Marg.$\quad$&$\quad$Cond.$\quad$& $\quad$Marg.$\quad$&$\quad$Cond.$\quad$& $\quad$Marg.$\quad$\\
			\hline 
			ESS/\#iter& 0.013 & 0.208 & 0.005& 0.076 & 0.006 & 0.142\\
			IAT& 81.266& 4.937 &215.0& 15.634 &126.80& 8.011\\
		\end{tabular}
	}
\end{table}

However,  one iteration of the marginal algorithm is associated with a higher computational cost mainly due to the evaluation of the integrals in $\mathcal{M}_{\text{Bern}}$ and $\mathcal{M}_{\text{NB}}$. In particular, the computation of $\mathcal{M}_{\text{NB}}$ requires approximating an infinite sum. Here, we employed a Monte Carlo approximation based on 100 samples. Alternatively,  we could rely on numerical approximation or introduce an auxiliary variable to be sampled from its full conditional.

 Additionally, the variability of the posterior distributions obtained with the marginal algorithm depends on the availability of an efficient mechanism for sampling the auxiliary variables corresponding to the inner mixtures $U_1,\ldots, U_M$ from their full conditional. Here, given the importance of $U_1,\ldots, U_M$ in determining the full conditional of the allocating variables  $\bm{c}$ and $\bm{z}$, we sample the auxiliary variables any time a new inner cluster is created, in addition to sampling them at the end of the \texttt{for} cycle as described in Section~3.

While the marginal algorithm can be used to derive a point estimate of  the predictive distribution for a new observation, it does not provide appropriate credible intervals for either outer or inner mixtures. Thus, in terms of uncertainty quantification, it can be used only for deriving credible balls for the clustering structure.

On the contrary, the conditional algorithm provides proper uncertainty quantification for both the mixtures and the clustering, while it can also be straightforwardly implemented.

\section*{ S4 Data} \label{ap:data}
\begin{figure}[!h]
	\caption{\label{fig:barplot}Bar plots of the frequencies for the $d=7$ responses across the seven days of the week. For visualisation purposes, plots contain only counts smaller than 20.}
	\includegraphics[width=0.5\textwidth]{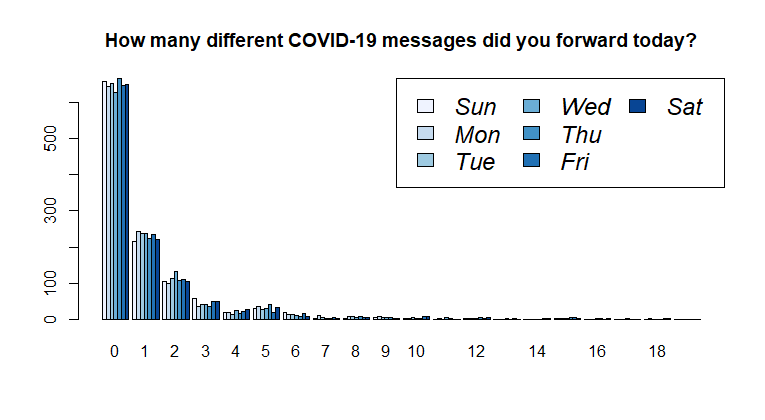}
	\includegraphics[width=0.5\textwidth]{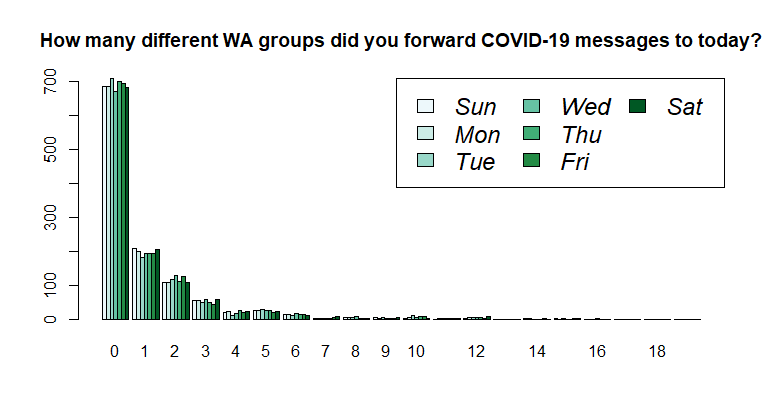}
\end{figure}
\begin{figure}[!h]
	\includegraphics[width=0.5\textwidth]{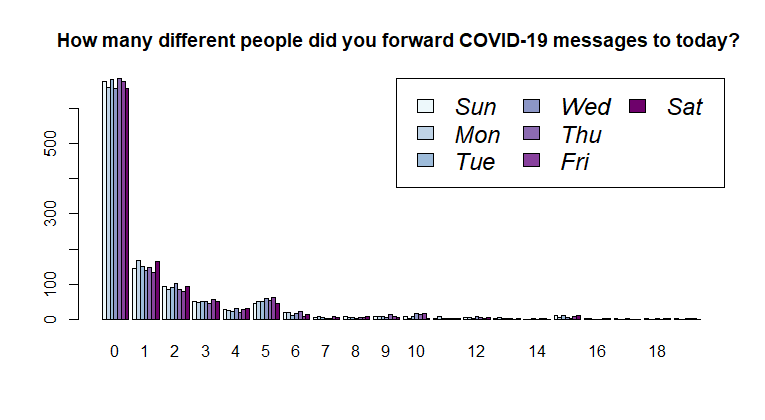}
	\includegraphics[width=0.5\textwidth]{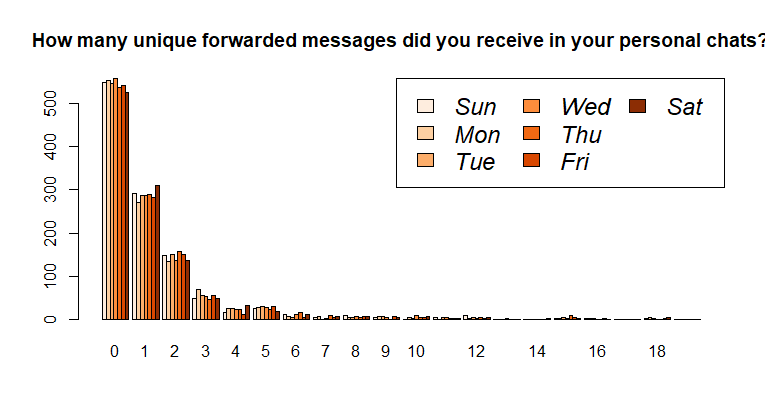}
	\includegraphics[width=0.5\textwidth]{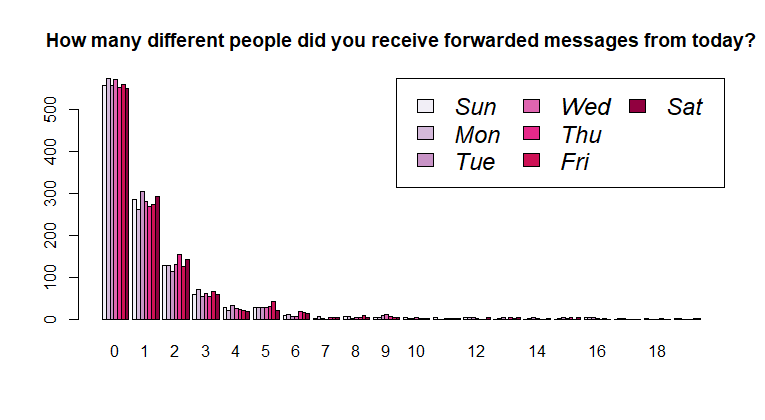}
	\includegraphics[width=0.5\textwidth]{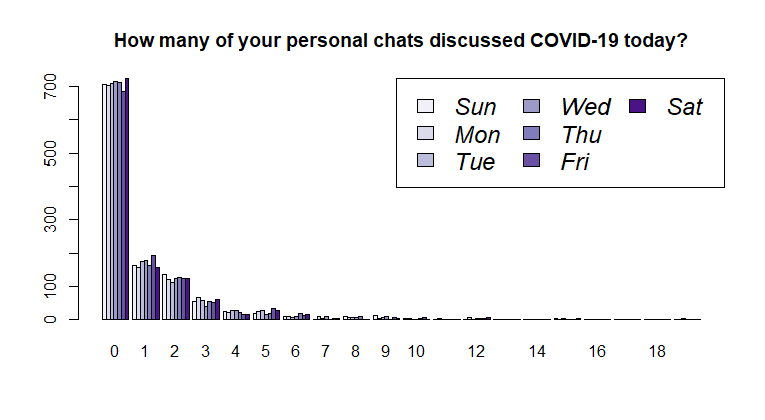}
	\includegraphics[width=0.5\textwidth]{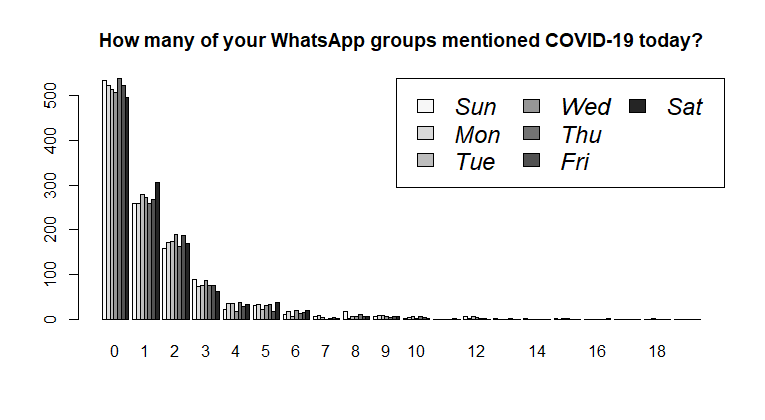}
\end{figure}
\end{changemargin}

\end{document}

%% file: figures/Figure_ZeroInflated.tikz
\begin{tikzpicture}
	
	\matrix[matrix of nodes, nodes in empty cells,
	row sep=-0.5\pgflinewidth,
	column sep=-0.5\pgflinewidth] (M1) {
		|[zero]| & |[zero]| & |[one]| & |[zero]| & |[one]| & |[zero]| & |[one]| \\
		|[zero]| & |[zero]| & |[one]| & |[one]| & |[zero]| & |[one]| & |[one]| \\
		|[one]| & |[one]| & |[zero]| & |[zero]| & |[one]| & |[zero]| & |[zero]| \\
	};
	\draw[dashed, path fading=south] (M1-3-1.south west) -- +(0,-0.75);
	\draw[dashed, path fading=south] ($(M1-3-1.south)!.5!(M1-3-2.south)$) -- +(0,-0.75);
	\draw[dashed, path fading=south] ($(M1-3-2.south)!.5!(M1-3-3.south)$) -- +(0,-0.75);
	\draw[dashed, path fading=south] ($(M1-3-3.south)!.5!(M1-3-4.south)$) -- +(0,-0.75);
	\draw[dashed, path fading=south] ($(M1-3-4.south)!.5!(M1-3-5.south)$) -- +(0,-0.75);
	\draw[dashed, path fading=south] ($(M1-3-5.south)!.5!(M1-3-6.south)$) -- +(0,-0.75);
	\draw[dashed, path fading=south] ($(M1-3-6.south)!.5!(M1-3-7.south)$) -- +(0,-0.75);
	\draw[dashed, path fading=south] (M1-3-7.south east) -- +(0,-0.75);
	\node[anchor=north west,inner sep=0] at ($(M1-1-7.north east)!.5!(M1-2-7.north east)$) {\includegraphics[width=0.1\textwidth]{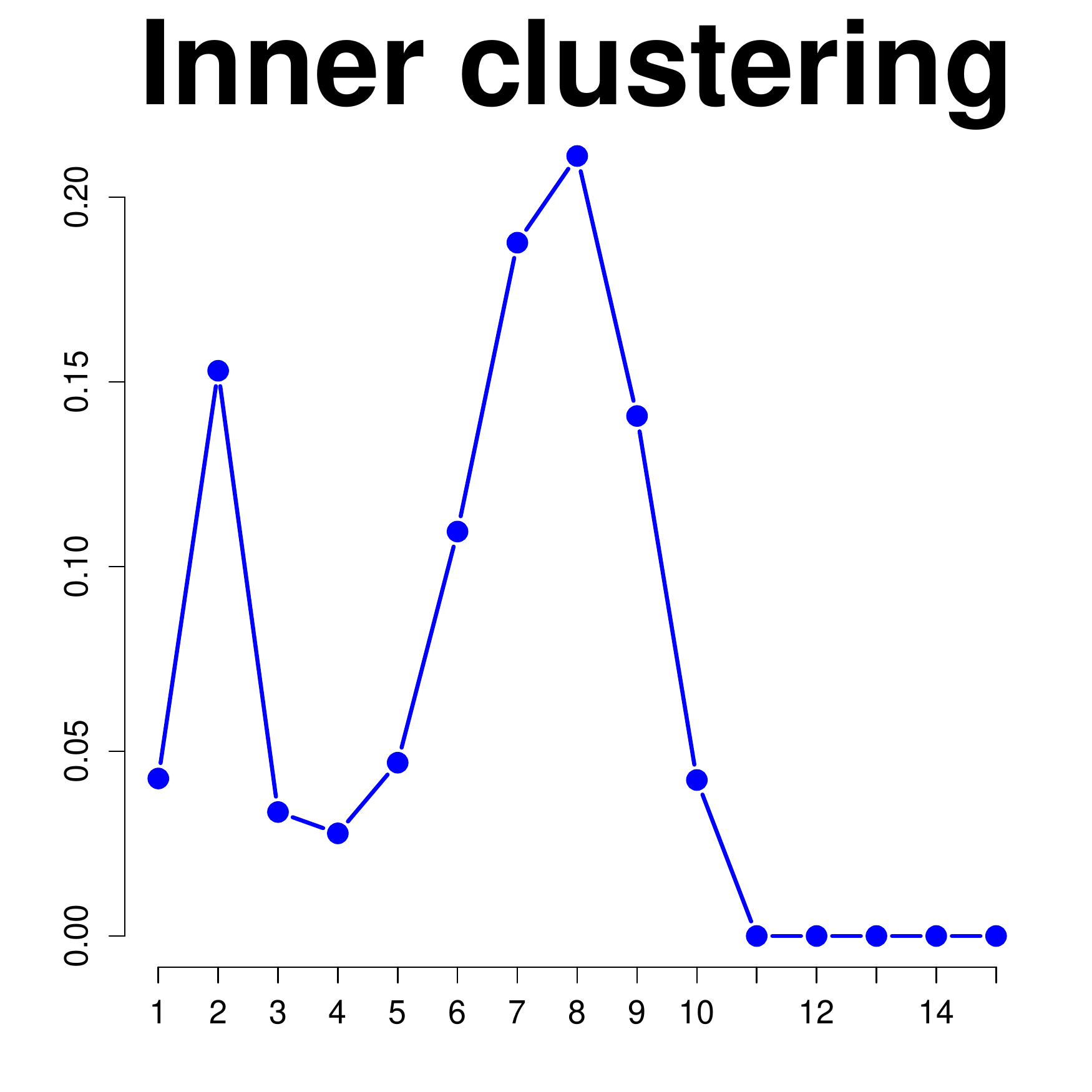}};
	\draw[draw={rgb,255:red,102; green,102; blue,255}, rounded corners, line width=0.65mm] (M1-1-1.north west)+(-0.1,0.15) rectangle +(+4.9,-2.4);
	\node[titlerect={\normalsize Outer Cluster 1}] at (M1-1-4.north) {};

	\matrix[matrix of nodes, nodes in empty cells,
	row sep=-0.5\pgflinewidth,
	column sep=-0.5\pgflinewidth] at (0,-3.5) (M2) {
		|[zero]| & |[zero]| & |[zero]| & |[one]| & |[zero]| & |[zero]| & |[zero]| \\
		|[one]| & |[one]| & |[zero]| & |[zero]| & |[zero]| & |[zero]| & |[zero]| \\
		|[zero]| & |[zero]| & |[zero]| & |[zero]| & |[zero]| & |[zero]| & |[zero]| \\
	};
	\draw[dashed, path fading=north] (M2-1-1.north west) -- +(0,0.75);
	\draw[dashed, path fading=north] ($(M2-1-1.north)!.5!(M2-1-2.north)$) -- +(0,0.75);
	\draw[dashed, path fading=north] ($(M2-1-2.north)!.5!(M2-1-3.north)$) -- +(0,0.75);
	\draw[dashed, path fading=north] ($(M2-1-3.north)!.5!(M2-1-4.north)$) -- +(0,0.75);
	\draw[dashed, path fading=north] ($(M2-1-4.north)!.5!(M2-1-5.north)$) -- +(0,0.75);
	\draw[dashed, path fading=north] ($(M2-1-5.north)!.5!(M2-1-6.north)$) -- +(0,0.75);
	\draw[dashed, path fading=north] ($(M2-1-6.north)!.5!(M2-1-7.north)$) -- +(0,0.75);
	\draw[dashed, path fading=north] (M2-1-7.north east) -- +(0,0.75);
	\draw[dashed, path fading=south] (M2-3-1.south west) -- +(0,-0.75);
	\draw[dashed, path fading=south] ($(M2-3-1.south)!.5!(M2-3-2.south)$) -- +(0,-0.75);
	\draw[dashed, path fading=south] ($(M2-3-2.south)!.5!(M2-3-3.south)$) -- +(0,-0.75);
	\draw[dashed, path fading=south] ($(M2-3-3.south)!.5!(M2-3-4.south)$) -- +(0,-0.75);
	\draw[dashed, path fading=south] ($(M2-3-4.south)!.5!(M2-3-5.south)$) -- +(0,-0.75);
	\draw[dashed, path fading=south] ($(M2-3-5.south)!.5!(M2-3-6.south)$) -- +(0,-0.75);
	\draw[dashed, path fading=south] ($(M2-3-6.south)!.5!(M2-3-7.south)$) -- +(0,-0.75);
	\draw[dashed, path fading=south] (M2-3-7.south east) -- +(0,-0.75);
	\node[anchor=north west,inner sep=0] at ($(M2-1-7.north east)!.5!(M2-2-7.north east)$) {\includegraphics[width=0.1\textwidth]{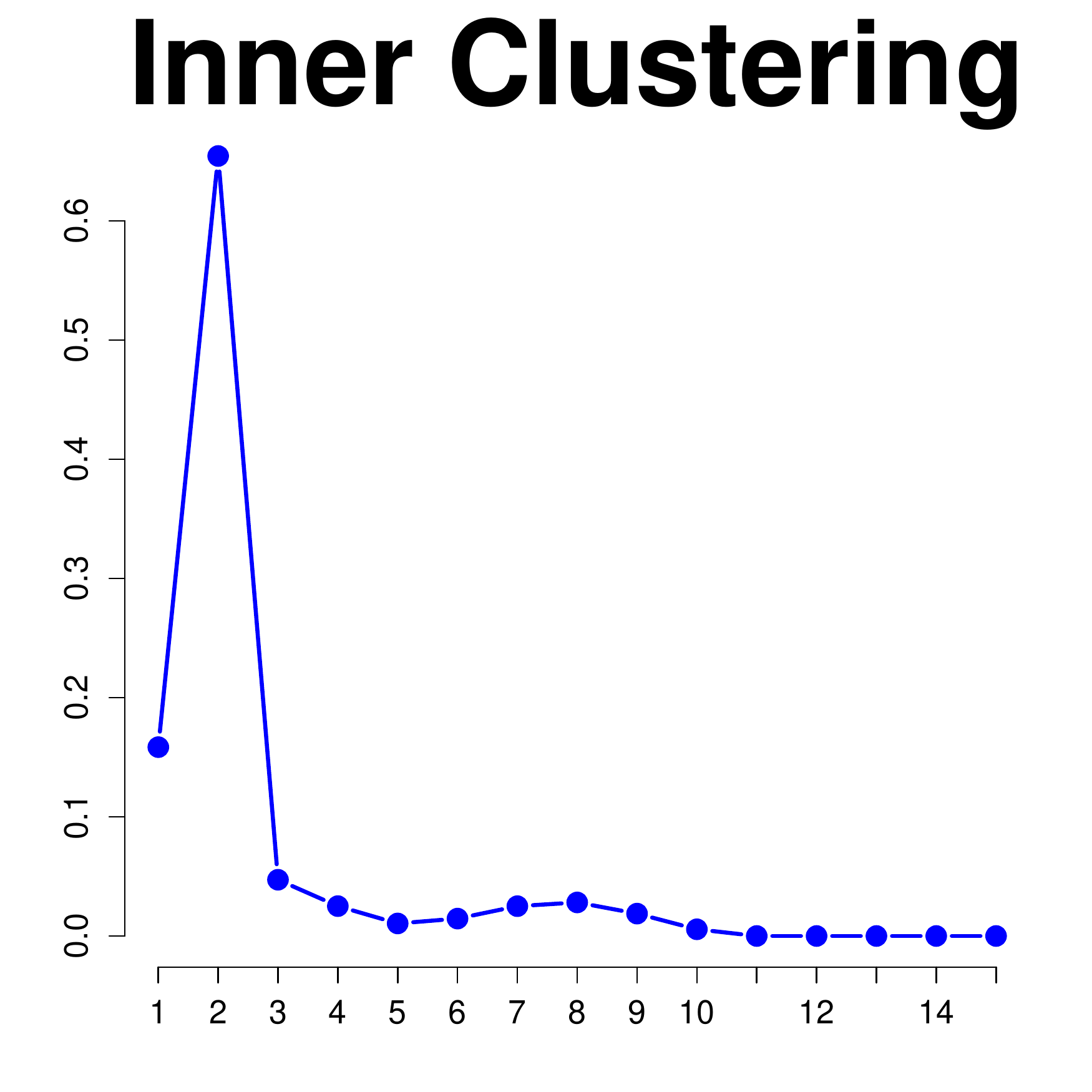}};
	\draw[draw={rgb,255:red,255; green,178; blue,102}, rounded corners, line width=0.65mm, dashed] (M2-1-1.north west)+(-0.1,0.85) rectangle +(+4.9,-2.4);
	\node[titlerect={\normalsize Outer Cluster 2}, above=6pt] at (M2-1-4.north) {};

	\matrix[matrix of nodes, nodes in empty cells,
	row sep=-0.5\pgflinewidth,
	column sep=-0.5\pgflinewidth] at (0,-7) (M3) {
		|[one]| & |[one]| & |[one]| & |[one]| & |[one]| & |[one]| & |[one]| \\
		|[one]| & |[one]| & |[zero]| & |[one]| & |[one]| & |[one]| & |[zero]| \\
		|[one]| & |[one]| & |[one]| & |[zero]| & |[one]| & |[one]| & |[one]| \\
	};
	\draw[dashed, path fading=north] (M3-1-1.north west) -- +(0,0.75);
	\draw[dashed, path fading=north] ($(M3-1-1.north)!.5!(M3-1-2.north)$) -- +(0,0.75);
	\draw[dashed, path fading=north] ($(M3-1-2.north)!.5!(M3-1-3.north)$) -- +(0,0.75);
	\draw[dashed, path fading=north] ($(M3-1-3.north)!.5!(M3-1-4.north)$) -- +(0,0.75);
	\draw[dashed, path fading=north] ($(M3-1-4.north)!.5!(M3-1-5.north)$) -- +(0,0.75);
	\draw[dashed, path fading=north] ($(M3-1-5.north)!.5!(M3-1-6.north)$) -- +(0,0.75);
	\draw[dashed, path fading=north] ($(M3-1-6.north)!.5!(M3-1-7.north)$) -- +(0,0.75);
	\draw[dashed, path fading=north] (M3-1-7.north east) -- +(0,0.75);
	\node[anchor=north west,inner sep=0] at ($(M3-1-7.north east)!.5!(M3-2-7.north east)$) {\includegraphics[width=0.1\textwidth]{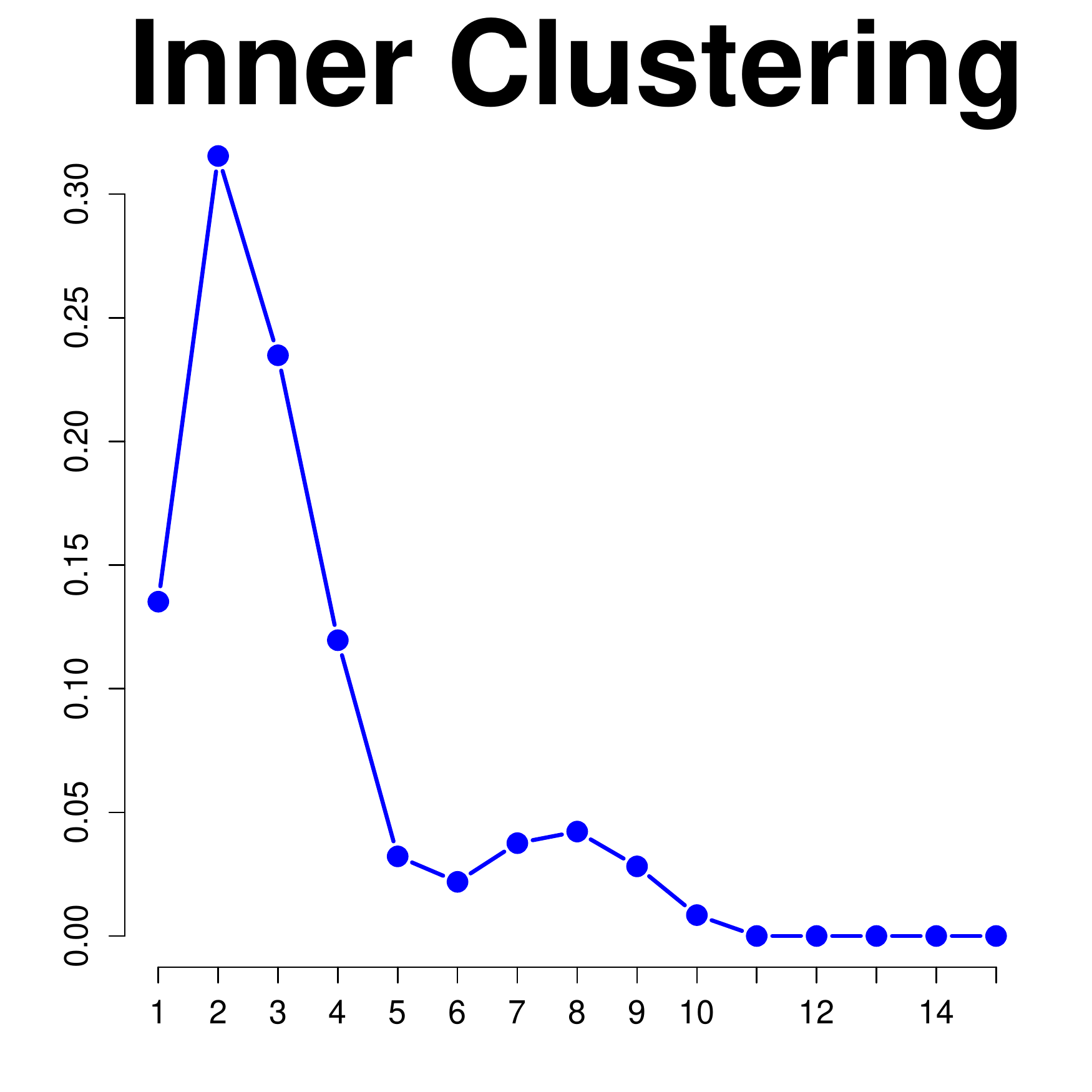}};
	\draw[draw={rgb,255:red,102; green,255; blue,178}, rounded corners, line width=0.65mm, dashdotted] (M3-1-1.north west)+(-0.1,0.85) rectangle +(+4.9,-1.625);
	\node[titlerect={\normalsize Outer Cluster 3},above=6pt] at (M3-1-4.north) {};
	
\end{tikzpicture}